\documentclass[twocolumn,10pt,aps,prd]{revtex4-2}
\usepackage{hyperref,graphicx,amsfonts,amsmath,amssymb}
\usepackage[latin1]{inputenc}
\usepackage{multirow}
\usepackage{subcaption}
\usepackage{diagbox}
\usepackage{titlesec}
\usepackage{soul}
\usepackage{xcolor}
\usepackage{cleveref}
\usepackage{siunitx}
\def\beq{\begin{equation}}
	\def\eeq{\end{equation}}
\def\beqa{\begin{eqnarray}}
	\def\eeqa{\end{eqnarray}}
\def\sba{s_{\beta-\alpha}}
\def\cba{c_{\beta-\alpha}}

\def\tb{t_{\beta}}

\def\m12{m_{12}}
\def\h125{h_{125}}
\def\ab{ab^{-1}}
\begin{document}
\title{Probing Mis-Aligned Type-I Two-Higgs-Doublet Model in $e^+e^- \to ZH$ at Future Lepton Colliders}
	
	\author{Majid Hashemi}
	\email{hashemi$_$mj@shirazu.ac.ir}
	\affiliation{Department of Physics, College of Science, Shiraz University, Shiraz, 71946-84795, Iran}

\begin{abstract}
A detailed collider study of Higgs production in the process
$e^+e^- \to ZH$ is presented within the framework of the Two-Higgs-Doublet Model (2HDM), with particular emphasis on mis-aligned scenarios in the Type-I realization.
The analysis is performed at a future $e^+e^-$ collider with
$\sqrt{s}=500$~GeV and an integrated luminosity of $10~\mathrm{ab}^{-1}$, considering heavy Higgs masses in the range $160 \leq m_H \leq 400$~GeV.
The analysis covers heavy CP-even Higgs states with non-fermionic decay modes
such as $H\to WW$ and $H\to hh$, and incorporates final states arising from
leptonic and invisible $Z$ boson decays. Fast parametrized detector-level
simulations are employed to evaluate selection efficiencies, signal--background
discrimination, and exclusion sensitivities. It is shown that values of $|c_{\beta-\alpha}|>0.1$ can be excluded at the $95\%$ confidence level over almost the entire range $1\leq\tan\beta\leq10$ in the $H\to WW$ scenario, while the $H\to hh$ channel provides lower sensitivity.

\end{abstract}

\maketitle
\section{Introduction}

The discovery of a Higgs boson with a mass of approximately 125~GeV at the
Large Hadron Collider (LHC) has established the Standard Model (SM) Higgs sector
as a remarkably successful description of electroweak symmetry breaking
\cite{ATLAS:2012yve,CMS:2012qbp}.
Precision measurements of the Higgs boson properties have so far shown
excellent agreement with SM predictions
\cite{ATLAS:2022vkf,CMS:2022dwd},
placing increasingly stringent constraints on possible extensions of the scalar
sector.

Nevertheless, the Higgs sector remains one of the least explored components
of the SM, and extended scalar frameworks continue to be strongly motivated
both theoretically and phenomenologically.

Among these extensions, the Two-Higgs-Doublet Model (2HDM) represents one of
the simplest and most extensively studied scenarios
\cite{Branco:2011iw,Gunion:2002zf}.
It arises naturally in a variety of well-motivated frameworks, including
supersymmetric models and scenarios addressing electroweak baryogenesis \cite{Branco:2011iw}.

The extended scalar spectrum of the 2HDM leads to rich collider phenomenology,
with additional neutral and charged Higgs bosons that can be directly searched
for at current and future experiments.

Over the past decade, extensive experimental searches for additional Higgs
bosons have been performed at the LHC.
These searches have placed strong bounds on large regions of the 2HDM parameter
space, particularly in scenarios where new scalar states exhibit sizable
couplings to fermions.

However, significant portions of the Type-I 2HDM parameter space remain weakly constrained. Current LHC searches exclude parts of the low-$\tan\beta$ region over the heavy Higgs mass range considered in this work, with the strongest limits arising from $H/A\to\tau\tau$ searches \cite{LHC8CMStautau,T4_1}. At moderate and large $\tan\beta$, however, all fermionic Yukawa couplings are suppressed in the Type-I model, reducing the dominant production cross sections. In addition, away from the alignment limit the heavy Higgs acquires couplings to electroweak gauge bosons, enhancing competing decay modes such as $H\to WW$, $ZZ$, and $hh$, thereby reducing the sensitivity of conventional fermionic search channels. The large Standard Model backgrounds, together with experimental systematic uncertainties in these bosonic final states, further limit the sensitivity of hadron colliders, leaving substantial regions of the parameter space unexplored.

Much of the existing literature has focused on the so-called alignment limit of the 2HDM, in which one of the CP-even Higgs mass eigenstates is aligned with the direction of the electroweak vacuum expectation value in Higgs-field space. In this limit, the aligned Higgs boson possesses the same tree-level couplings to gauge bosons and fermions as the Standard Model Higgs boson. In the conventional parameterization, this corresponds to $\cos(\beta-\alpha)=0$, where $\beta$ and $\alpha$ denote the mixing angles of the scalar sector. Consequently, the observed 125 GeV Higgs boson becomes SM-like. 

While current Higgs signal-strength measurements strongly constrain deviations from the alignment limit, as demonstrated by LHC data analyses \cite{LHC8,LHC9,hb3,hb4}, they still permit controlled but phenomenologically relevant departures from alignment \cite{Carena:2013ooa,Dev:2014yca,globalfit}, particularly when correlations among Higgs observables and theoretical constraints are taken into account. Throughout this work, the term \emph{mis-alignment} refers to these non-zero deviations from the alignment limit, i.e. $\cos(\beta-\alpha)\neq0$, which modify the Higgs couplings and open additional production and decay channels absent or highly suppressed in the exact alignment limit.

Interestingly, in certain realizations of the 2HDM, small but non-negligible
departures from alignment can even lead to improved agreement with Higgs data.

In this context, the Type-I 2HDM is of particular interest.
In contrast to other Yukawa realizations, all fermions couple to a single Higgs doublet, resulting in a characteristic and universal modification of the fermionic Higgs couplings away from the alignment limit \cite{Ferreira:2012my,Craig:2013hca}.
As a result, mis-aligned Type-I scenarios can evade many of the stringent bounds
from direct searches and flavor observables, while simultaneously enhancing
Higgs self-interactions and couplings to electroweak gauge bosons.
This opens up the possibility of observing Higgs production and decay channels
that are either absent or highly suppressed in the alignment limit.

The exploration of such scenarios is especially well suited to future
high-energy and high-luminosity $e^+e^-$ colliders.
Lepton colliders offer a clean experimental environment, precisely known
initial states, and excellent reconstruction capabilities, allowing for
model-independent Higgs studies and detailed investigations of extended scalar
sectors \cite{Baer:2013cma,ILC:2013jhg,cliccdr,clichiggs1,clichiggs2,cepchiggs,lepton2HDM}.

Recent studies have investigated the potential of future $e^+e^-$ colliders to
perform precision Higgs coupling measurements and to probe extended scalar sectors
in a variety of models \cite{Fuji,fmaltoni,Hou}.
Additional work has specifically explored heavy neutral scalar production channels
and their sensitivity at high-energy lepton colliders \cite{Robens:2021}.

The $e^+e^- \rightarrow ZH$ provides a unique opportunity to study the Higgs boson in a model-independent manner through the recoil-mass technique, in which the Higgs boson is identified from the four-momentum of the associated $Z$ boson without explicitly reconstructing its decay products. This method was originally proposed in the context of Higgs production at lepton colliders~\cite{Bjorken1977} and has since become one of the benchmark measurements for future Higgs factories ~\cite{GarciaAbia2000,Marshall2016,MI}. 

The objective of the present work is different. Rather than employing the recoil-mass method, the feasibility of directly reconstructing the heavy CP-even Higgs boson from its visible decay products is investigated. This approach exploits the characteristic event kinematics of the reconstructed Higgs candidate to enhance the discrimination between signal and Standard Model backgrounds, thereby providing a complementary strategy to recoil-based analyses.

Motivated by these considerations, a dedicated collider-level study
of mis-aligned Type-I 2HDM scenarios at future $e^+e^-$ colliders is performed.
Higgs production in association with a $Z$ boson is considered, exploiting both
visible and invisible $Z$ decay modes.

Special emphasis is placed on Higgs decay topologies characteristic of
mis-alignment, including decays into pairs of electroweak gauge bosons as well
as Higgs-to-Higgs decays.

By analyzing multiple final states and reconstruction strategies, the discovery
potential and background rejection capabilities are assessed for a
broad class of phenomenologically viable Type-I 2HDM scenarios.
\section{Theoretical Framework}

\subsection{The Two-Higgs-Doublet Model}

The Two-Higgs-Doublet Model (2HDM) extends the SM scalar sector by introducing
a second $SU(2)_L$ scalar doublet with hypercharge $Y=1$,
\begin{equation}
	\Phi_i =
	\begin{pmatrix}
		\phi_i^+ \\
		\frac{1}{\sqrt{2}}(v_i + \rho_i + i \eta_i)
	\end{pmatrix},
	\qquad i = 1,2 ,
\end{equation}
where $v_1$ and $v_2$ denote the vacuum expectation values (VEVs) of the two
doublets, satisfying $v^2 = v_1^2 + v_2^2 = (246~\mathrm{GeV})^2$.

The scalar sector of the 2HDM considered in this work is based on a
CP-conserving scalar potential with a softly broken $\mathbb{Z}_2$
symmetry, introduced to prevent tree-level flavor-changing neutral
currents. Under this symmetry,
$\Phi_1\rightarrow\Phi_1$ and
$\Phi_2\rightarrow-\Phi_2$, which forbids the
$\lambda_6$ and $\lambda_7$ quartic interactions. The symmetry is
softly broken by the dimension-two parameter $m_{12}^2$. The resulting
scalar potential is given by~\cite{Branco:2011iw,Gunion:2002zf}:

\begin{align}
	V(\Phi_1,\Phi_2) ={}&
	m_{11}^2 \Phi_1^\dagger \Phi_1
	+ m_{22}^2 \Phi_2^\dagger \Phi_2
	- m_{12}^2 \left( \Phi_1^\dagger \Phi_2 + \mathrm{h.c.} \right) \nonumber\\
	&+ \frac{\lambda_1}{2} (\Phi_1^\dagger \Phi_1)^2
	+ \frac{\lambda_2}{2} (\Phi_2^\dagger \Phi_2)^2
	\nonumber\\
	&+ \lambda_3 (\Phi_1^\dagger \Phi_1)(\Phi_2^\dagger \Phi_2) + \lambda_4 (\Phi_1^\dagger \Phi_2)(\Phi_2^\dagger \Phi_1)
	\nonumber\\
	&+ \frac{\lambda_5}{2} \left[ (\Phi_1^\dagger \Phi_2)^2 + \mathrm{h.c.} \right] .
\end{align}

After electroweak symmetry breaking, the physical Higgs spectrum consists of
two CP-even neutral scalars ($h$ and $H$), one CP-odd scalar ($A$), and a pair
of charged Higgs bosons ($H^\pm$).
The CP-even mass eigenstates are obtained by rotating the weak eigenstates
$(\rho_1,\rho_2)$ via a mixing angle $\alpha$,
$\begin{pmatrix}
	H \\
	h
\end{pmatrix}
=
\begin{pmatrix}
	\cos\alpha & \sin\alpha \\
	-\sin\alpha & \cos\alpha
\end{pmatrix}$
$\begin{pmatrix}
	\rho_1 \\
	\rho_2
\end{pmatrix}.$

The ratio of the VEVs is defined as
\begin{equation}
\tb = \frac{v_2}{v_1},
\end{equation}
which, together with the angle $\alpha$, governs the couplings of the Higgs
bosons to fermions and gauge bosons. The usual abbreviations like $\tan\beta\equiv \tb$ and $\cos(\beta-\alpha)\equiv \cba$ are used throughout the paper.

The couplings of the CP-even Higgs states to electroweak gauge bosons are
given by \cite{Gunion:2002zf,Carena:2013ooa}
\begin{equation}
	g_{hVV} = \sba\, g_{hVV}^{\rm SM}, \qquad
	g_{HVV} = \cba\, g_{hVV}^{\rm SM},
\end{equation}
where $V=W,Z$ and $g_{hVV}^{\rm SM}=2m_V^2/v$ denotes the corresponding
Standard Model Higgs coupling. At the alignment limit, i.e., $\cba\to0$, the light CP-even Higgs $h$ acquires SM-like couplings,
while the heavy Higgs couplings to electroweak gauge bosons vanish.

The trilinear Higgs self-couplings, such as $g_{Hhh}$, depend on the masses,
mixing angles and quartic couplings of the scalar potential, and therefore do
not exhibit a simple universal scaling with $\cba$.

\subsection{Mis-Alignment in Type-I 2HDM}

In the Type-I realization of the 2HDM, all fermions couple exclusively to one
Higgs doublet, conventionally chosen to be $\Phi_2$.
As a consequence, the Yukawa couplings of both CP-even Higgs bosons to fermions take the form
\begin{equation}
	g_{hff} = \frac{\cos\alpha}{\sin\beta}g_{hff}^{\rm SM}, \qquad
	g_{Hff} = \frac{\sin\alpha}{\sin\beta}g_{hff}^{\rm SM}.
\end{equation}
where $g_{hff}^{\rm SM}=m_f/v$.

Away from the alignment limit, these couplings can be significantly suppressed,
especially at moderate to large values of $\tan\beta$.
At the same time, the coupling of the heavy CP-even Higgs boson to gauge bosons
remains proportional to $\cba$ and can be sizable in mis-aligned
scenarios \cite{Ferreira:2012my,Chen:2013rba}.
This characteristic coupling pattern allows the heavy Higgs boson to evade
stringent fermion-based searches while exhibiting enhanced bosonic decay modes.

In addition, mis-alignment can lead to large trilinear Higgs self-couplings,
most notably the $Hhh$ coupling, which governs the decay
\begin{equation}
	H \to hh ,
\end{equation}
together with the gauge-boson-mediated decays
\begin{equation}
	H \to WW, \qquad H \to ZZ ,
\end{equation}
which dominate the heavy Higgs phenomenology over wide regions of the
Type-I 2HDM parameter space \cite{Djouadi:2005gj,Baglio:2014nea}.

For this analysis, a set of publicly available tools are employed. Theoretical constraints and the decay properties of neutral Higgs bosons are obtained using \texttt{2HDMC-1.8.0} \cite{2hdmc1,2hdm2,2hdm3}, while \texttt{HiggsTools-1} \cite{higgstools} provides experimental data from LEP, Tevatron, and LHC measurements. Event generation and cross-section calculations for the lepton collider are performed with \texttt{WHIZARD-3.1.2} \cite{whizard1,whizard2}, incorporating beam spectra through \texttt{circe2} \cite{circe2}. \texttt{PYTHIA-8.3.09} \cite{pythia} is used to simulate multiparticle interactions, final state radiation, and parton showering. Detector effects are modeled via \texttt{Delphes-3.5.0} \cite{delphes1,delphes2,delphes3}, utilizing the \texttt{ILCgen} card for reconstructing physical objects. Visualization of the results is carried out with Python3 libraries \texttt{NumPy} \cite{numpy} and \texttt{Matplotlib} \cite{matplotlib}.

To quantify the allowed level of mis-alignment in light of current Higgs data,
a global fit is performed using \texttt{HiggsSignals}.

\begin{figure}[h!] \centering \includegraphics[width=0.45\textwidth]{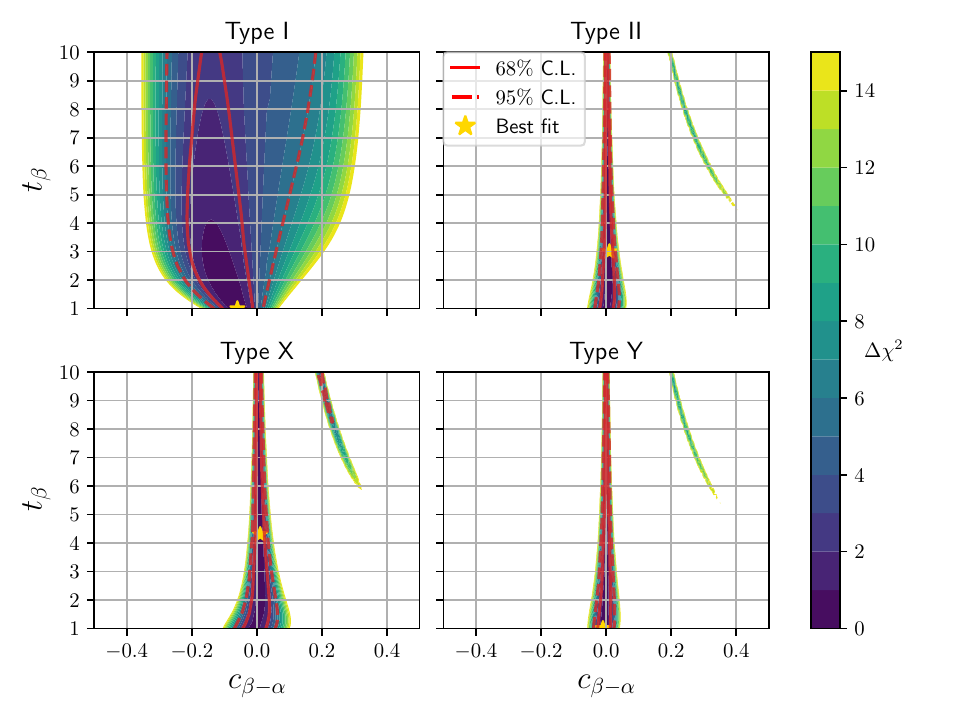} \caption{$\Delta\chi^2$ distributions in the
		$(\cba,\tb)$ plane obtained with \texttt{HiggsSignals}
		for the four 2HDM types. Solid (dashed) curves indicate the $68\%$ ($95\%$)
		confidence regions. The yellow star marks the best-fit point.}
	\label{fig:DeltaChi2} \end{figure}

\Cref{fig:DeltaChi2} shows the $\Delta\chi^2 \equiv \chi^2-\chi^2_{\min}$
distributions in the $(\cba,\tb)$ plane for the four
Yukawa realizations of the 2HDM. The solid and dashed contours correspond to
the $68\%$ and $95\%$ confidence levels, respectively, while the yellow star
denotes the best-fit point in each scenario.
The numerical values of the best-fit coordinates,
together with the minimum $\chi^2$, are summarized in
Table~\ref{tab:HiggsSignals}.

The Type-I realization yields a lower minimum $\chi^2$ than other types. However, this difference should not be interpreted as a statistical significance. Since the two Yukawa realizations represent distinct hypotheses, the $\Delta\chi^2$ values serve only as a relative measure of the goodness of fit.
\begin{table}[h!] 
	\centering 
	\caption{Best-fit $\chi^2$ values and corresponding
		$(\cba,\tb)$ coordinates from \texttt{HiggsSignals}
		for the four 2HDM types.}
	
	 \label{tab:HiggsSignals} 
	 \begin{tabular}{cccc} 
	 	\hline 
	 	\textbf{Type} & $\chi^2_\text{min}$ & $\cba$ & $\tb$ \\ \hline I & 149.002 & -0.06 & 1.0 \\
	 	 II & 152.012 & 0.01 & 3.0 \\
	 	   X & 151.863 & 0.01 & 4.3 \\
	 	  Y & 152.269 & -0.01 & 1.0 \\
	 	    \hline 
 	    \end{tabular} 
     \end{table}
     
The best fit position reflects the combined constraints from all Higgs signal-strength
measurements included in the fit and should not be interpreted as being driven
by a single experimental channel. Since the Higgs coupling modifiers depend
linearly on both $c_{\beta-\alpha}$ and $\tan\beta$, the likelihood is
generally not symmetric under the transformation
$c_{\beta-\alpha}\rightarrow -c_{\beta-\alpha}$, resulting in a mildly
asymmetric allowed region.

The fit indicates that controlled departures from alignment remain fully consistent with current Higgs signal-strength measurements and can even provide a comparable or slightly improved description of the data. Consequently, the heavy CP-even Higgs boson can retain suppressed
fermionic couplings but appreciable couplings to gauge bosons and Higgs pairs,
thereby enhancing the $H \to WW$, $ZZ$, and $hh$ decay modes. These bosonic
channels therefore constitute the primary targets of our collider analysis.

\section{Heavy Neutral Higgs in the Type-I 2HDM}

\subsection{Theoretical constraints and exclusions}
Perturbativity of the scalar sector requires the quartic couplings $\lambda_i$ to remain moderate, conventionally $|\lambda_i| \lesssim 4\pi$, while the Yukawa couplings satisfy $|y_f| \lesssim \sqrt{4\pi}$ \cite{Branco2012}. Tree-level unitarity of $2\to2$ scattering of Higgs and longitudinal gauge bosons imposes bounds on combinations of $\lambda_i$ (the Lee--Quigg--Thacker eigenvalues) such that $|\Lambda| \le 8\pi$. Vacuum stability demands that the scalar potential is bounded from below, leading to the conditions
\begin{align}
	\lambda_1 &> 0, & \lambda_2 &> 0, \nonumber\\
	\lambda_3 + \sqrt{\lambda_1 \lambda_2} &> 0, & 
	\lambda_3 + \lambda_4 - |\lambda_5| + \sqrt{\lambda_1 \lambda_2} &> 0.
\end{align}
In practice, these translate into upper limits on the Higgs masses and splittings: large mass differences between heavy scalars drive the quartic couplings to large values, potentially violating unitarity or stability. In the Type-I 2HDM, one typically requires $|\lambda_i| \lesssim \mathcal{O}(1\text{--}10)$ and small mass splittings to satisfy these conditions \cite{Chen2019}.

\begin{figure}[h!]
	\centering
	\includegraphics[width=0.48\textwidth]{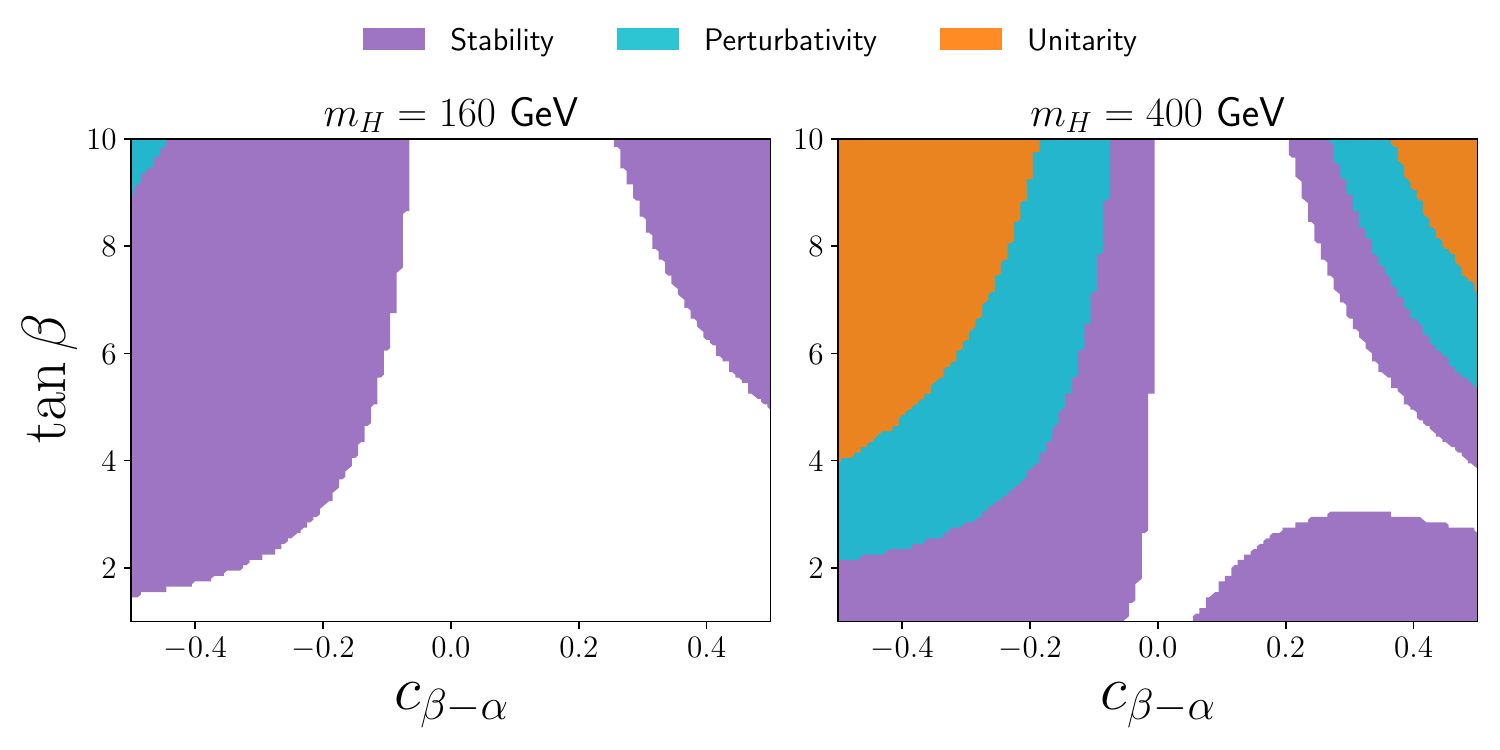}
\caption{Theoretically excluded regions in the $(\tan\beta,\,c_{\beta-\alpha})$
	plane for $m_H=160$ and $400~\mathrm{GeV}$. Different colors indicate the
	theoretical constraints responsible for excluding the corresponding parameter
	points.}
	\label{fig:theory_excl}
\end{figure}

These theoretical requirements also constrain the $(\tb, \cba)$ parameter space. Regions that violate any of these conditions are considered unphysical and are therefore excluded \cite{Branco2012}. \Cref{fig:theory_excl} illustrates the excluded regions for the lowest and highest Higgs masses, $m_H = 160$ and $400~\mathrm{GeV}$. The left-hand region with $c_{\beta-\alpha} < 0$ is almost entirely excluded by vacuum stability requirement for all masses, whereas the right-hand region ($c_{\beta-\alpha} > 0$) is excluded only at low and high $\tb$, leaving a moderate $\tb$ corridor open. This pattern reflects the interplay between the scalar potential parameters and the theoretical conditions.

These theoretical exclusions are essential when interpreting branching ratios and $\sigma \times \mathrm{BR}$ results. In particular, the moderate $\tan\beta$ region for positive $c_{\beta-\alpha}$ remains a viable window for heavy Higgs phenomenology, compatible with theoretical constraints and observable signals in the $H \to WW$ and $H \to hh$ channels \cite{CMS:2018lkl,c2016}.
\subsection{Flavor physics bounds}
In the Type-I 2HDM, all fermion couplings scale as $\cot\beta$, so flavor
observables provide the strongest constraints at low $\tan\beta$.
In particular, the radiative decay $B \to X_s \gamma$, which receives
charged-Higgs--top loop contributions, imposes stringent bounds on the
charged-Higgs sector and disfavors very small $\tan\beta$ values for
$M_{H^\pm}\sim\mathcal{O}(200$--$500~\mathrm{GeV})$
\cite{Misiak,Misiak2}. Other flavor observables, such as
$B_s\to\mu^+\mu^-$, generally provide weaker or complementary constraints in
the Type-I 2HDM over the parameter region considered here \cite{Arbey:2017gmh,CMSLHCbATLASBsMuMu}.
The flavor constraints are complementary to the
direct heavy-Higgs searches discussed below, which further restrict the
low-$\tan\beta$ region through LHC data.
\subsection{Experimental constraints from direct searches}
In addition to the theoretical requirements discussed above, the parameter space of the model is constrained by direct searches for additional Higgs bosons at the LHC. These searches probe resonant production of heavy scalar states through their decays into gauge bosons and fermions, placing significant constraints on the $(m_H,\tb)$ parameter space.

\begin{figure}[h!]
	\centering
	\includegraphics[width=0.45\textwidth]{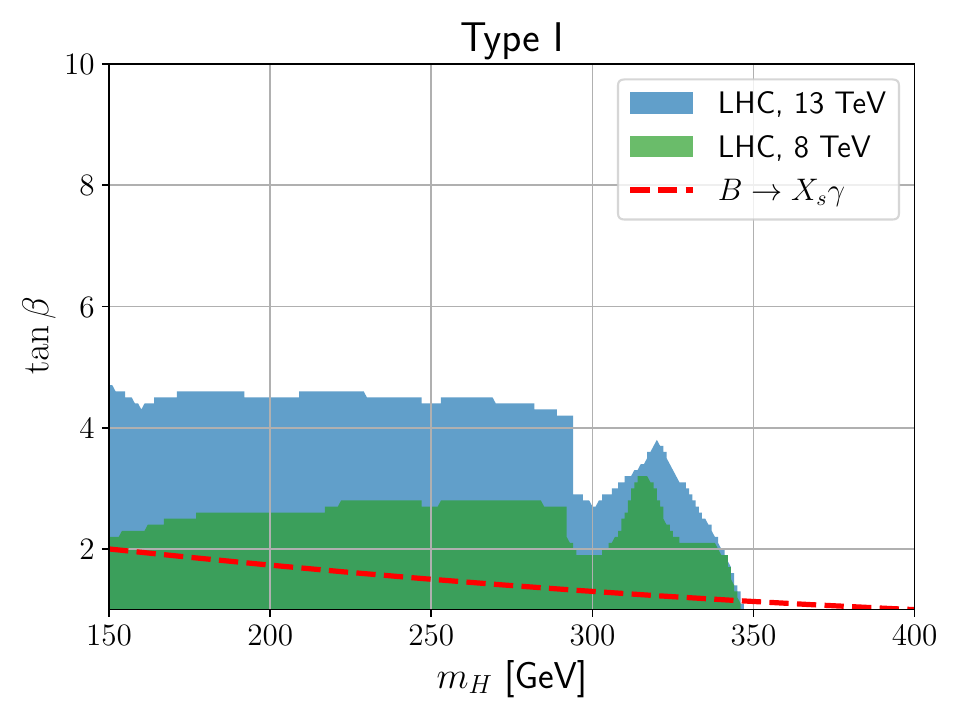}
\caption{Experimental exclusion limits in the $(m_H,\tan\beta)$ plane for the
	Type-I 2HDM from LHC heavy Higgs searches at $\sqrt{s}=8$ and
	$13~\mathrm{TeV}$. Shaded regions are excluded at $95\%$ confidence level.
	The dashed curve indicates the approximate lower bound from
	$B\to X_s\gamma$ assuming $m_{H^\pm}=m_H$ \cite{Misiak2}.}
	\label{fig:LHC_exclusions}
\end{figure}

\Cref{fig:LHC_exclusions} illustrates the current exclusion limits for the
Type-I 2HDM obtained from LHC data at $\sqrt{s}=8$ and
$13~\mathrm{TeV}$ in the alignment limit. The shaded regions indicate
parameter points excluded at the $95\%$ confidence level, with the strongest
sensitivity arising from $H\to\tau\tau$ searches at 8~TeV
\cite{LHC8CMStautau} and 13~TeV \cite{T4_1}. As expected for the Type-I 2HDM,
the constraints are strongest at low $\tb$, where the Yukawa couplings are
less suppressed and the production cross sections are correspondingly
enhanced. For comparison, the approximate lower bound on $\tan\beta$ inferred from
$B\to X_s\gamma$ in the Type-I 2HDM is also shown as a dashed curve,
translated to the present benchmark under the assumption
$m_{H^\pm}=m_H$ \cite{Misiak2}.

Although the collider analysis presented in this work is performed for
non-zero values of $\cos(\beta-\alpha)$, the experimental exclusions shown in
Fig.~\ref{fig:LHC_exclusions} are presented in the alignment limit and are
included primarily as a reference to illustrate the experimentally favored
regions of the $(m_H,\tan\beta)$ parameter space. Moderate deviations from the
alignment limit modify the precise exclusion contours through changes in the
heavy Higgs production rates and branching fractions, but the overall pattern
of stronger constraints at low $\tan\beta$ remains qualitatively unchanged.

\subsection{Electroweak precision measurements}
Oblique parameters $S$, $T$, and $U$ strongly constrain the scalar spectrum. The $T$ parameter, sensitive to custodial symmetry breaking, typically dominates:
\begin{equation}
	\Delta T \sim F(m_{H^\pm}^2, m_H^2) + F(m_{H^\pm}^2, m_A^2) - F(m_H^2, m_A^2) + \dots,
\end{equation}
where $F$ is an $\mathcal{O}(1)$ function of mass-squared differences. Precision data require $\Delta T$ to be small, enforcing approximate mass degeneracy: either $m_H \simeq m_{H^\pm}$ or $m_A \simeq m_{H^\pm}$. Large splittings $\gg v$ are disfavored \cite{Chen2019}. The $S$ parameter provides weaker constraints but also prefers moderate splittings. Overall, EW precision measurements demand a near-degenerate scalar spectrum or custodial-symmetric alignment.

\subsection{Higgs coupling fits (125~GeV Higgs)}
LHC measurements of the 125~GeV Higgs impose stringent bounds on the mixing angle $\cba$ and indirectly on heavy Higgs properties. In the alignment limit ($\cba=0$), the 125~GeV Higgs couplings are SM-like, while $H$ and $A$ have suppressed gauge interactions. Deviations from alignment induce coupling shifts; global fits (CMS 36~fb$^{-1}$, 13~TeV) allow only a narrow band around $\cba=0$ at 95\% CL \cite{CMS:2018lkl}. Moderate deviations $|\cba|$ are largely excluded for moderate $\tb$, whereas small deviations are consistent with data. Future precision (HL-LHC or Higgs factories) may tighten this to $|\cba| \lesssim 0.05$ \cite{c2016,Chen2019}.  

Direct searches for additional Higgs bosons at ATLAS/CMS provide complementary bounds. Channels such as $H/A \to \tau\tau$, $H \to ZZ/WW$, or $H \to hh$ exclude very low $\tb$ ($\lesssim 1$) for $m_H \sim 200$--500~GeV. Combining collider, Higgs, and flavor constraints indicates that a Type-I heavy Higgs in this mass range must lie close to alignment ($|\cba| \ll 1$) and satisfy $\tb \gtrsim 1$--2.

\subsection{Benchmark Parameterization}

For the numerical analysis, the focus is made on experimentally viable mis-aligned
configurations of the Type-I 2HDM that satisfy all current theoretical,
flavor, electroweak, and Higgs-coupling constraints discussed above.
Rather than restricting the analysis to the exact alignment limit
($c_{\beta-\alpha}=0$), small but finite deviations are probed,
which remain allowed by present data and lead to enhanced bosonic production
and decay rates of the heavy CP-even Higgs boson. These controlled departures
from alignment constitute the primary target of this study.

The Type-I 2HDM parameter space is described in terms
of physical Higgs masses, mixing angles, and the soft $\mathbb{Z}_2$-breaking
parameter.
Motivated by electroweak precision constraints and the latest measurements of
the $W$ boson mass, a degenerate heavy Higgs spectrum is adopted,
\begin{equation}
	m_H = m_A = m_{H^\pm},
\end{equation}
which preserves custodial symmetry and suppresses the contributions to the
electroweak oblique parameters, in particular the $\Delta\rho$ ($T$)
parameter \cite{Peskin:1990zt,Peskin:1991sw,Grimus:2008nb,drho1,drho2,drho3}.
With this assumption, the model is parametrized by the reduced set
\begin{equation}
	\left\{
	m_h,\,
	m_H,\,
	\tb,\,
	\cba,\,
	m_{12}^2
	\right\},
\end{equation}
where the light CP-even Higgs boson $h$ is identified with the observed
125~GeV state, and $m_H$ denotes the common mass of the heavy scalar states.
The parameter scan is performed over
\begin{equation}
	\begin{aligned}
		160 &\le m_H \le 400~\mathrm{GeV},\\
		1 &\le \tan\beta \le 10,\\
		-0.5 &\le c_{\beta-\alpha} \le 0.5.
	\end{aligned}
\end{equation}
while $m_A$ and $m_{H^\pm}$ are fixed by the degeneracy relation above and are
therefore not scanned independently.

\begin{figure}[h!]
	\centering
	\includegraphics[width=0.49\textwidth,height=0.2\textheight]{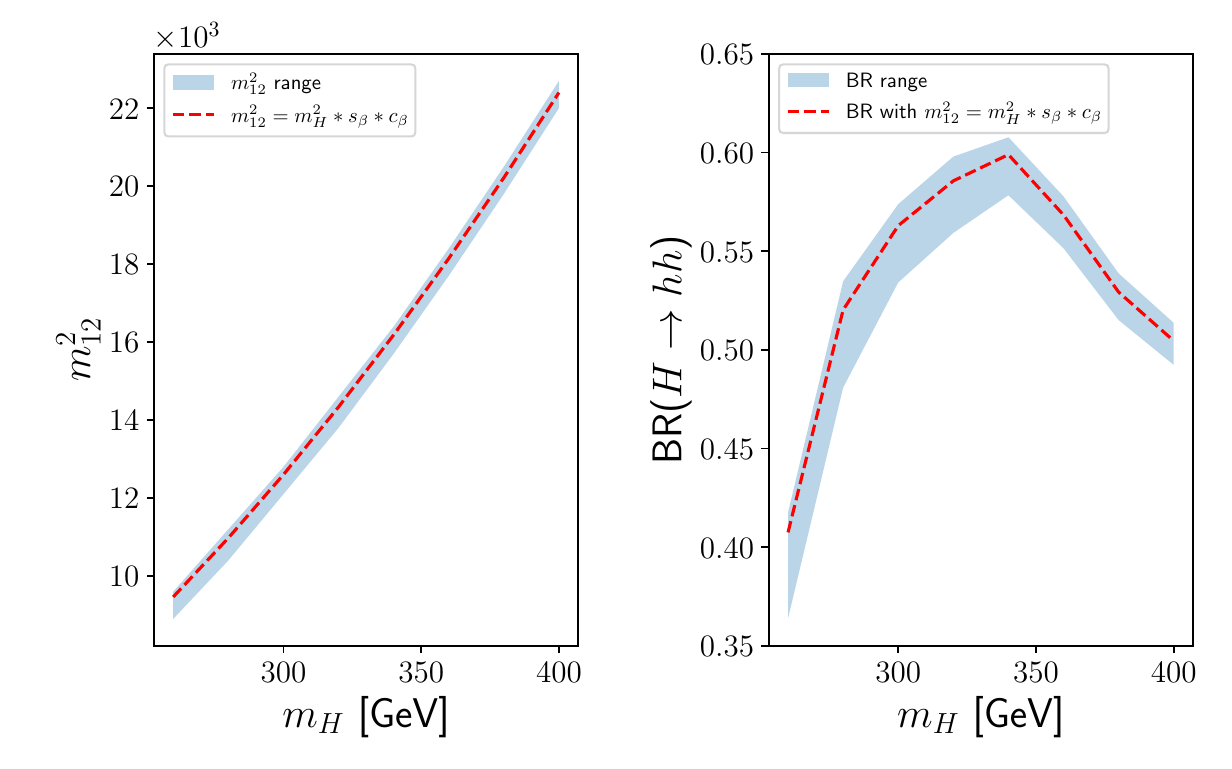}
\caption{Left: Theoretically allowed values of $m_{12}^2$ as a function of the
	heavy Higgs mass together with the benchmark choice
	$m_{12}^2=m_H^2\sin\beta\cos\beta$. Right: Corresponding variation of
	$\mathrm{BR}(H\to hh)$ over the allowed $m_{12}^2$ range.}
	\label{fig:m12}
\end{figure}

The soft-breaking parameter is chosen as
\begin{equation}
	m_{12}^2=m_H^2\sin\beta\cos\beta,
\end{equation}
which represents a commonly adopted benchmark in phenomenological studies of the
2HDM \cite{decoupling}. Alternative values satisfying
theoretical constraints are also possible. Figure~\ref{fig:m12} illustrates the
theoretically allowed range of $m_{12}^2$ together with the adopted benchmark
choice. The corresponding variation of $\mathrm{BR}(H\to hh)$ is found to be
below approximately $10\%$ over the allowed parameter space, indicating that
the benchmark choice is representative and does not qualitatively affect the exclusion limits presented in this work.

All benchmark points satisfy theoretical consistency conditions, including
vacuum stability, perturbative unitarity, and the existence of a global
electroweak minimum, as well as current experimental constraints, prior to
event generation.

\section{Signal and Background Processes}

The process
\[
e^+ e^- \to ZH
\]
is studied at a center-of-mass energy relevant for future lepton colliders.
Both leptonic ($Z \to \ell^+\ell^-$) and invisible ($Z \to \nu\bar{\nu}$) decay
modes of the $Z$ boson are considered.

\begin{figure}[hbt!]
	\centering
	\begin{subfigure}{.24\textwidth}
		\includegraphics[width=0.9\linewidth,height=0.8\linewidth]{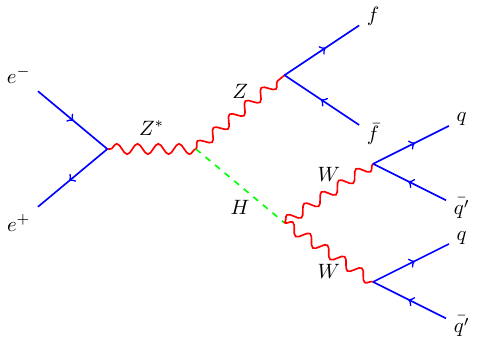}
		\caption{}
		\label{}
	\end{subfigure}%
	\begin{subfigure}{.24\textwidth}
		\includegraphics[width=0.9\linewidth,height=0.8\linewidth]{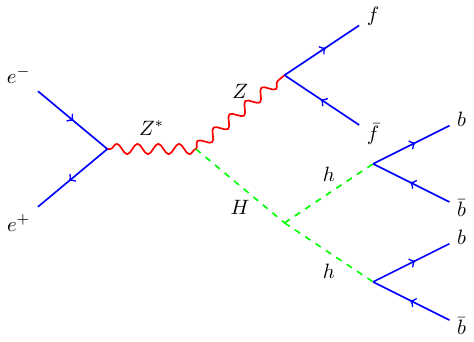}
		\caption{}
		\label{}
	\end{subfigure}
	\caption{The signal diagrams in two modes of the Higgs boson decay, $H \to WW$ (a) and $H \to hh$ (b).}
	\label{fig:H_decay_diagrams}
\end{figure}

For light Higgs masses, the decay
\begin{equation}
	H \to WW \to q\bar{q} q\bar{q},
\end{equation}
is considered, where both $W$ bosons decay hadronically. For heavier Higgs masses, the dominant channel
\begin{equation}
	H \to hh \to b\bar{b}b\bar{b}
\end{equation}
is investigated. The corresponding Feynman diagrams for these processes are shown in \cref{fig:H_decay_diagrams}.

The dominant SM backgrounds arise from $t\bar{t}$ and $t\bar{t}b\bar{b}$
production. In what follows, the light Higgs search is performed in the 4$j$ final state while the heavy Higgs is searched through the 4$b$-j final state. The $\ell^+\ell^-$ and $\nu\bar{\nu}$ channels denote the associated $Z$ boson decay modes.

\section{Collider Energy Choice}

\begin{figure}[h!]
	\centering
	\includegraphics[width=0.47\textwidth]{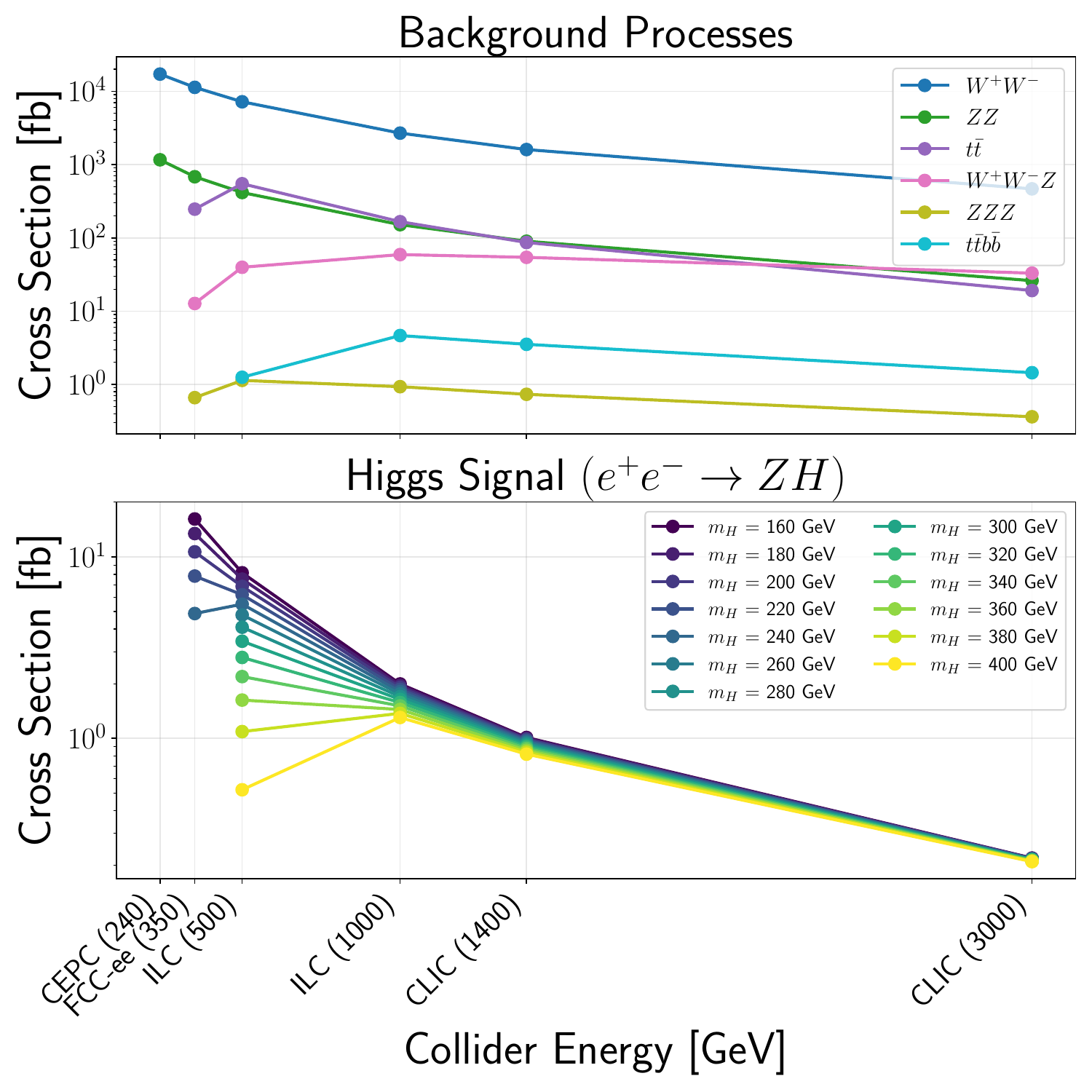}
	\caption{Cross sections for the background processes and the signal with different Higgs boson masses as a function of the collider center of mass energy. Vertical lines indicate the center-of-mass energies of the CEPC (240 GeV), FCC-ee (350 GeV), ILC (500 GeV and 1000 GeV), and CLIC (1400 GeV and 3000 GeV) proposals.}
	\label{fig:cross_sections}
\end{figure}

The search for heavy Higgs bosons in the $ZH$ production channel requires a center-of-mass energy sufficiently above the $m_H + m_Z$ threshold. The cross sections for the signal and background processes are shown in Figure~\ref{fig:cross_sections}. As expected, the signal cross sections generally decrease with collider energy, while the background contributions remain relatively stable.

\begin{figure}[h!]
	\centering
	\includegraphics[width=0.47\textwidth]{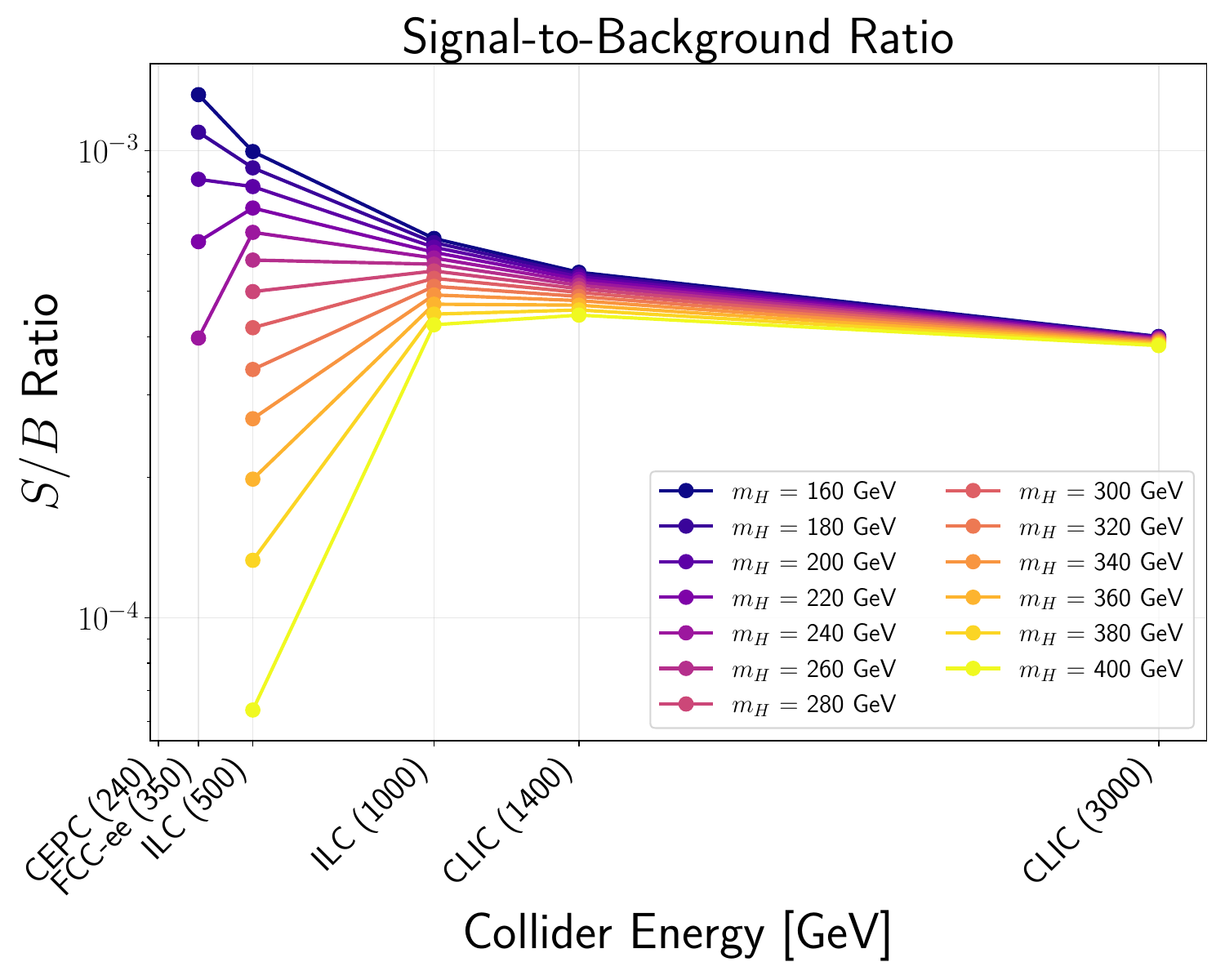}
	\caption{Signal-to-background ratio ($S/B$) as a function of collider energy for $e^+e^- \to ZH$ production, for Higgs boson masses ranging from 160 to 400 GeV.}
	\label{fig:s_over_b}
\end{figure}

The corresponding signal-to-background ratio ($S/B$) as a function of the collider energy is presented in Figure~\ref{fig:s_over_b} for Higgs masses from 160 to 400 GeV. The $S/B$ ratio improves with increasing energy for higher masses but then remains almost stable.

The ILC at 500 GeV is chosen as the baseline for this analysis. It is the first operating scenario capable of producing the entire benchmark mass range (160--400 GeV), while lower-energy colliders such as CEPC and FCC-ee lack sufficient kinematic reach. Higher-energy options like ILC-1000 or CLIC offer improved sensitivity for the highest masses but require substantially more time for realization. 

Although the HL-LHC is expected to considerably extend the sensitivity to additional Higgs bosons with an integrated luminosity of about $3~\mathrm{ab}^{-1}$, a future $e^+e^-$ collider provides a complementary environment with well-defined initial-state kinematics and substantially lower backgrounds. The present work therefore focuses on the 500~GeV ILC scenario, which offers an excellent balance between kinematic reach and sensitivity for the benchmark mass range considered.

\section{Cross Sections, Branching Ratios, and $\sigma \times \mathrm{BR}$}

\begin{figure}[h]
	\centering
	\includegraphics[width=0.45\textwidth]{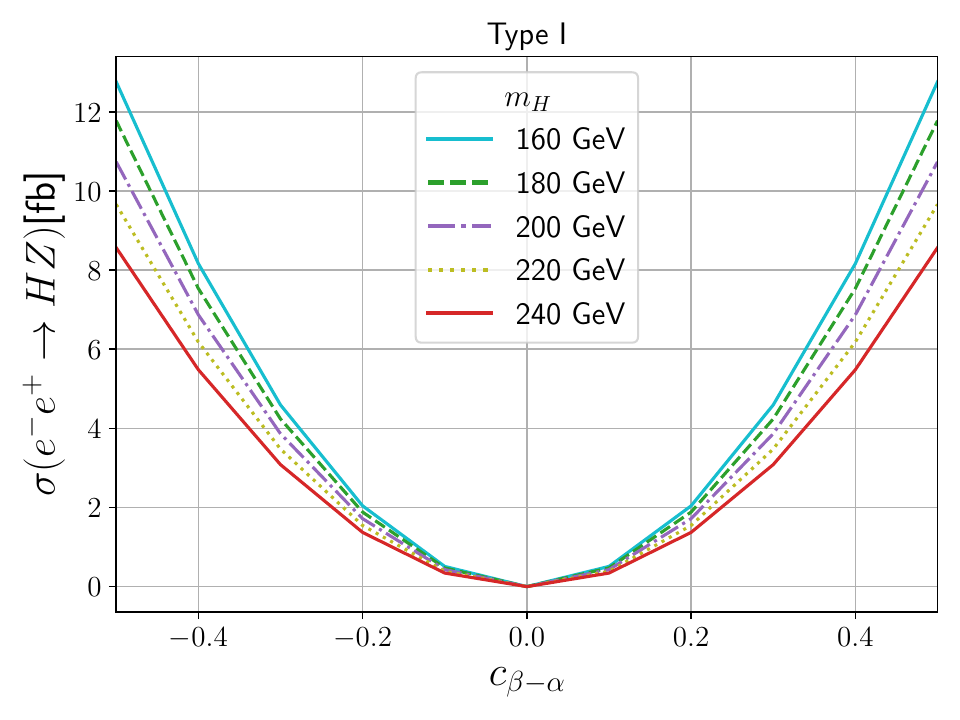}
	\caption{Production cross section $\sigma(e^+e^- \to HZ)$ for representative masses in the diboson-dominated regime.}
	\label{fig:xsec_light}
\end{figure}

\begin{figure}[h]
	\centering
	\includegraphics[width=0.45\textwidth]{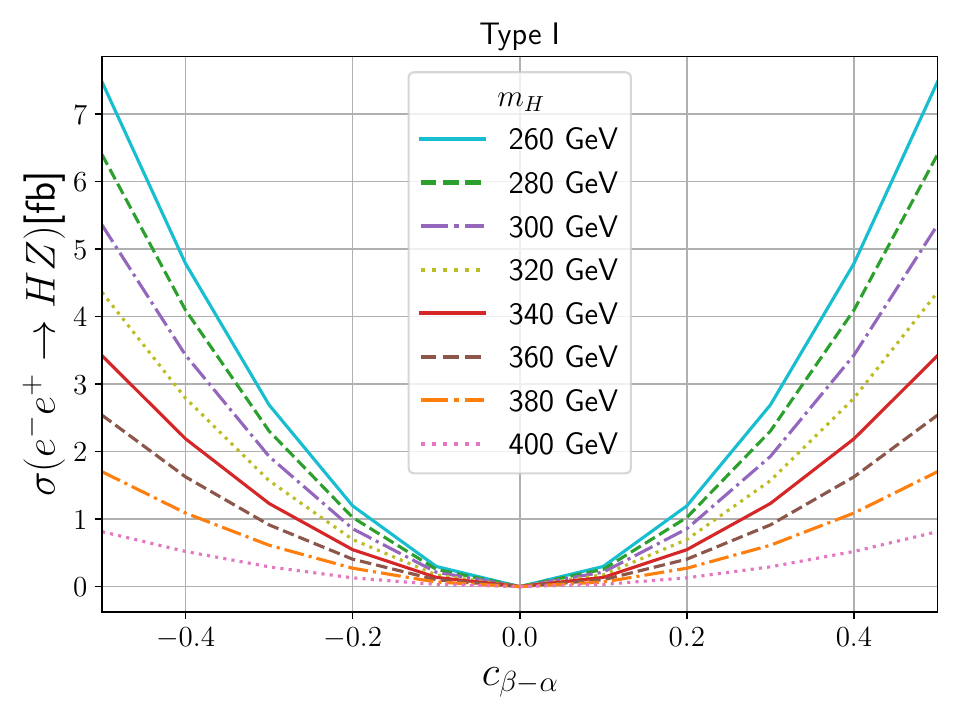}
	\caption{Production cross section $\sigma(e^+e^- \to HZ)$ for representative masses in the di-Higgs-dominated regime.}
	\label{fig:xsec_heavy}
\end{figure}

The production and decay of the heavy CP-even Higgs boson $H$ can be organized according to its dominant decay phenomenology. \Cref{fig:xsec_light,fig:xsec_heavy} show the production cross section for $e^+e^- \to HZ$ as a function of $\cba$ for phenomenologically representative masses consistent with the allowed parameter space
in the diboson-dominated ($H\to WW/ZZ$) and di-Higgs-dominated ($H\to hh$) regimes, respectively. The cross section scales as $\sigma(HZ)\propto \cba^2$, vanishing in the alignment limit and increasing away from it, with a reduction for heavier Higgs masses due to phase-space suppression.

\begin{figure}[h!]
	\centering
	\includegraphics[width=0.45\textwidth]{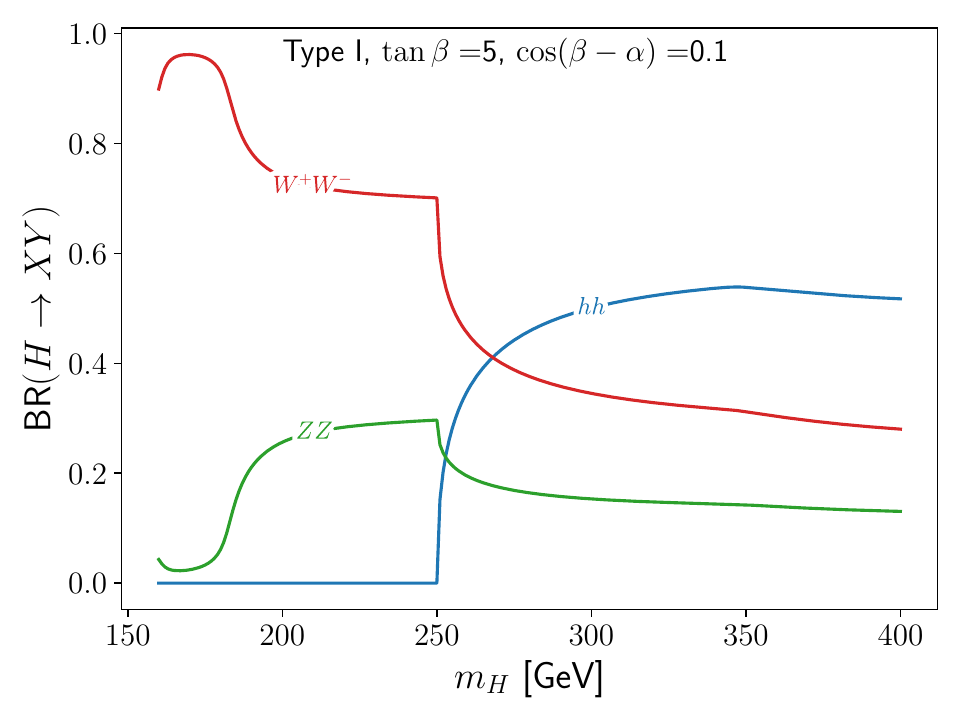}
	\caption{Branching ratio of Higgs boson decays for specific values of $\tb=5$ and $\cba=0.1$}
	\label{fig:BR_H_tb5}
\end{figure}

\begin{figure}[h!]
	\centering
	\includegraphics[width=0.45\textwidth]{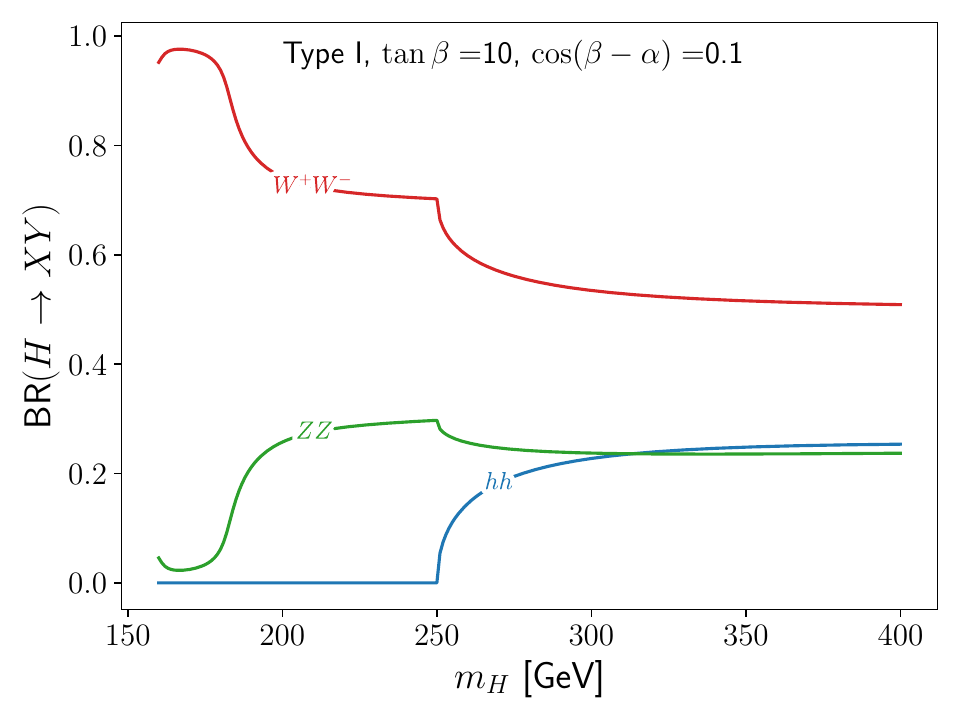}
	\caption{Branching ratio of Higgs boson decays for specific values of $\tb=10$ and $\cba=0.1$}
	\label{fig:BR_H_tb10}
\end{figure}

The corresponding branching ratios for the main decay channels are shown in \cref{fig:BR_H_tb5,fig:BR_H_tb10} for two values of $\tb=5$ and 10 at $\cba=0.1$. Comparing the two plots suggests that at moderate $\tb$ the $H\to WW$ will dominate until $H\to hh$ is kinematically turned on. For higher values of $\tb$ (e.g. $\tb=10$) $H\to WW$ is dominant in the whole region of heavy Higgs boson masses. 

\begin{figure}[h!]
	\centering
	\includegraphics[width=0.45\textwidth]{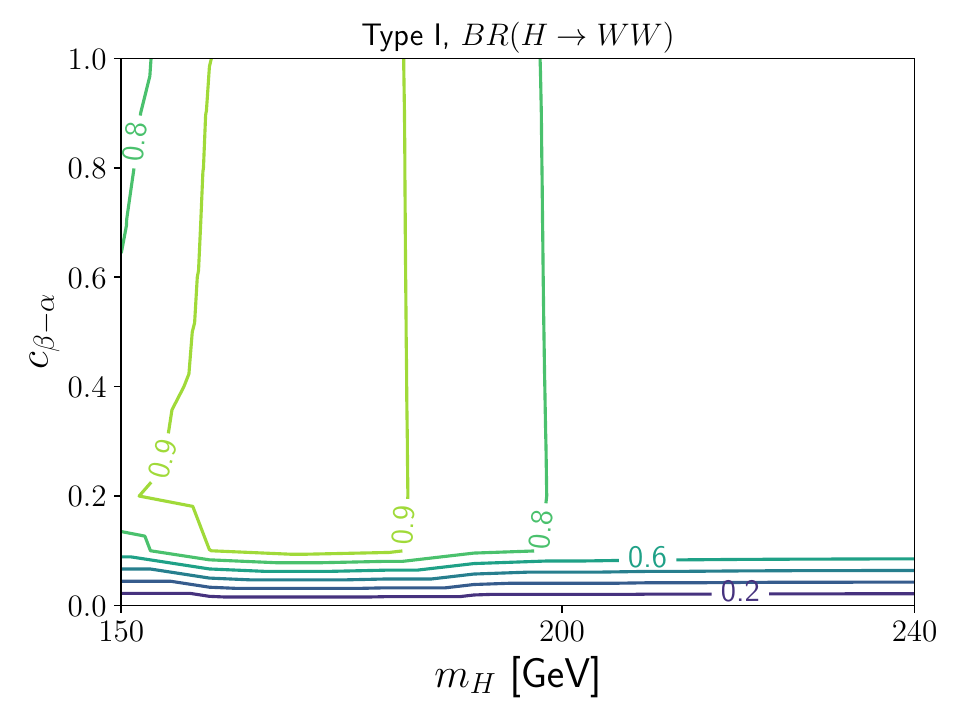}
	\caption{Branching ratio of $H \to WW$ in the $(c_{\beta-\alpha}, m_H)$ plane at $\tb = 5$. The branching ratio increases with $\cba$, reflecting the dependence of the HVV coupling on the departure from the alignment limit.}
	\label{fig:BR_H_WW}
\end{figure}

\begin{figure}[h!]
	\centering
	\includegraphics[width=0.45\textwidth]{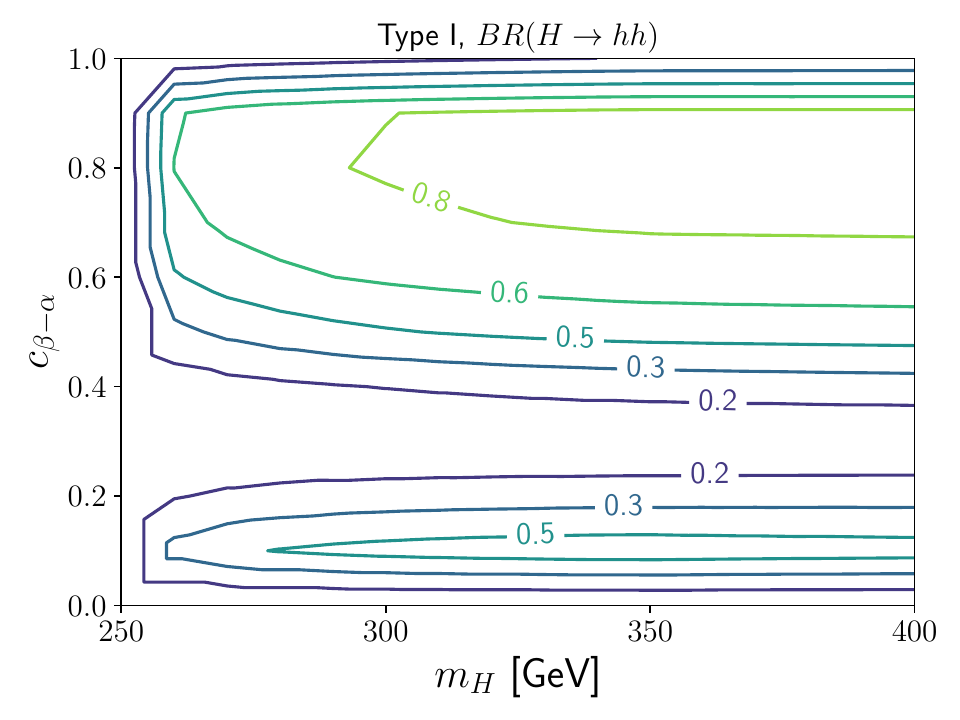}
	\caption{Branching ratio of $H \to hh$ in the $(c_{\beta-\alpha}, m_H)$ plane at $\tb = 5$. The valley around $c_{\beta-\alpha} \simeq 0.3$ arises from the partial cancellation between the linear and cubic $c_{\beta-\alpha}$ terms in the $Hhh$ coupling.}
	\label{fig:BR_H_hh}
\end{figure}

\Cref{fig:BR_H_WW,fig:BR_H_hh} show the corresponding branching ratios of $H\to WW$ and $H\to hh$ as a function of $\cba$. Using the full tree-level expression given in \cite{align1}
(see in particular Eqs.~(62) and (B.9)), one may expand the coupling around
alignment to obtain
\begin{equation}
	g_{Hhh}
	= A\,c_{\beta-\alpha}
	+ B\,c_{\beta-\alpha}^{3}
	+ \mathcal{O}(c_{\beta-\alpha}^{5}),
\end{equation}
where the coefficients $A$ and $B$ depend on the scalar masses and quartic
couplings of the potential.
For $m_H \gtrsim 250~\mathrm{GeV}$, the linear term is negative, while the cubic term is positive. The competition between these terms leads to a partial cancellation at an intermediate value, $c_{\beta-\alpha} \simeq 0.3$, suppressing the partial width $\Gamma(H \to hh)$ and producing the valley observed in the branching ratio. At larger $c_{\beta-\alpha}$, the cubic term dominates and the branching ratio rises again \cite{Arco:2022xum}.


\begin{figure}[h!]
	\centering
	\includegraphics[width=0.45\textwidth]{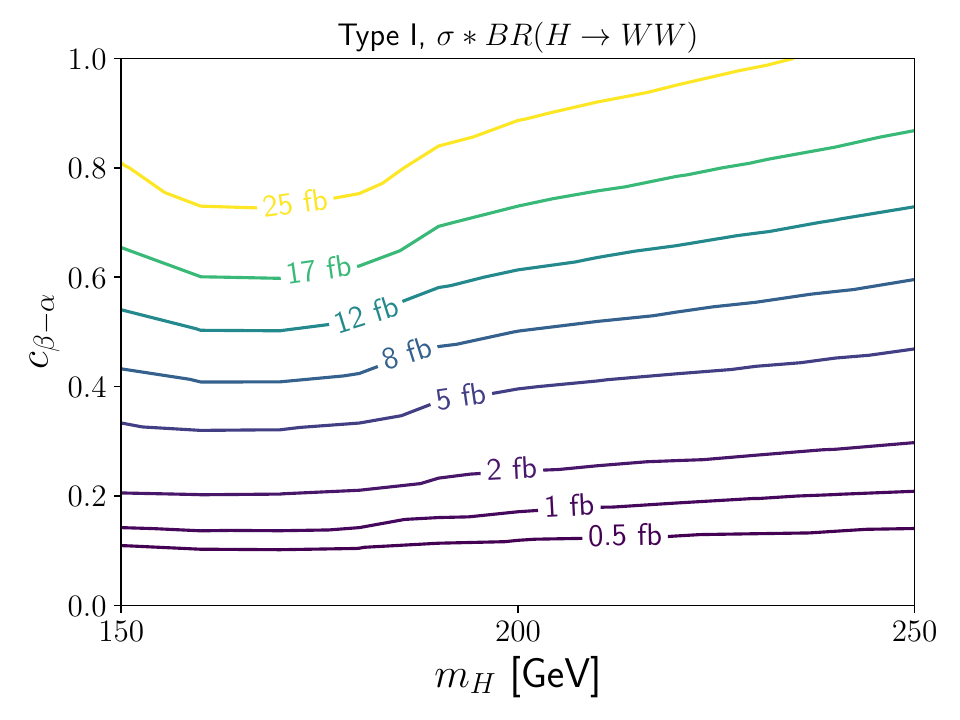}
	\caption{Production cross section times branching ratio, $\sigma \times \mathrm{BR}$, for $H \to WW$ in the $(c_{\beta-\alpha}, m_H)$ plane at $\tb = 5$.}
	\label{fig:SxBR_H_WW}
\end{figure}

\begin{figure}[h!]
	\centering
	\includegraphics[width=0.45\textwidth]{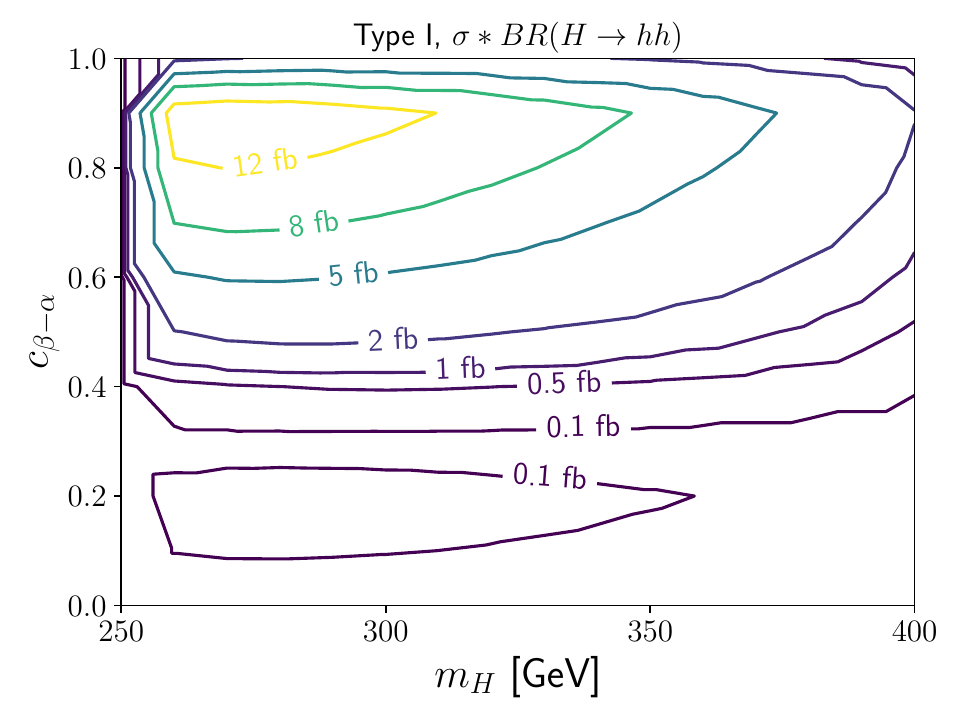}
	\caption{Production cross section times branching ratio, $\sigma \times \mathrm{BR}$, for $H \to hh$ in the $(c_{\beta-\alpha}, m_H)$ plane at $\tb = 5$. Suppression around $c_{\beta-\alpha} \simeq 0.3$ arises from the partial cancellation between the linear and cubic terms in the $Hhh$ coupling.}
	\label{fig:SxBR_H_hh}
\end{figure}

\Cref{fig:SxBR_H_WW,fig:SxBR_H_hh} combine production and decay, showing $\sigma \times \mathrm{BR}$ for the two dominant channels. The $H \to WW$ channel exhibits smooth behavior across the parameter space, while $H \to hh$ clearly reflects the nontrivial structure caused by the triple Higgs vertex. The suppression around $c_{\beta-\alpha} \simeq 0.3$ in both BR and $\sigma \times \mathrm{BR}$ is entirely due to the interference between the linear and cubic terms, and not from competition with other decay channels, as fermionic decays are suppressed at this $\tb$.

\section{Event Simulation and Analysis Strategy}

Signal and background events are generated at parton level and passed through
a full detector simulation.
Jets are reconstructed using standard clustering algorithms, and object
identification follows realistic detector performance assumptions.

Event selections are optimized separately for leptonic and invisible $Z$
decay channels.
They include requirements on jet multiplicities, $b$-tagging, missing energy,
and invariant mass reconstruction of intermediate resonances.

A cut--based analysis is performed targeting associated Higgs production
\begin{equation}
	e^+e^- \to ZH ,
\end{equation}
considering both leptonic and invisible decays of the $Z$ boson and two hadronic decay modes of the Higgs boson:
\begin{align}
	H \to WW \to 4j, \qquad
	H \to hh \to 4b .
\end{align}

\subsection{Object reconstruction and definitions}

All reconstructed objects are required to satisfy the fiducial acceptance
\begin{equation}
	p_T > 10~\mathrm{GeV}, \qquad |\eta| < 5 ,
\end{equation}
where $p_T$ is the transverse momentum and $\eta$ is the pseudorapidity. This acceptance applies to all jets and leptons unless otherwise stated.

\begin{itemize}
	\item \textbf{Jets:} Jets are reconstructed from final-state particles using the anti-$k_T$ algorithm \cite{antikt} with a distance parameter $R = 0.4$. Only jets passing the fiducial acceptance are considered. For the $H \to WW \to 4j$ channels, a light-jet selection based on anti-$b$-tagging is applied to reject $b$-jets, significantly reducing contributions from $t\bar{t}$ and $t\bar{t}b\bar{b}$ processes.
	
	\item \textbf{$b$-jets:} Jets are classified as $b$-tagged based on the $b$-tagging algorithm working point corresponding to an approximate efficiency of 90\%. Only $b$-tagged jets enter the $h \to b\bar{b}$ reconstruction in the $H \to hh$ channels where at least four $b$-tagged jets are required.
	
	\item \textbf{Leptons:} Electrons and muons (denoted as $\ell$) are required to be isolated from other activity. At least two leptons passing the fiducial acceptance are required for channels with $Z \to \ell^+\ell^-$. For the $Z \to \nu\bar{\nu}$ channels, a lepton veto is applied.
	
	\item \textbf{Missing transverse energy ($E_T^{miss}$):} The missing transverse momentum, $\vec{E}_T^{\rm miss}$, is reconstructed from all visible objects. The specific $E_T^{miss}$ requirements depend on the final state and are detailed in the respective sections below.
\end{itemize}


For the $H \to WW$ channels, several kinematic observables are constructed from the reconstructed Higgs candidate, obtained from the two reconstructed $W$ bosons, to enhance the separation between signal and Standard Model backgrounds.

\begin{itemize}
	
	\item \textbf{$\Delta R_{WW}$:}
	The angular separation between the two reconstructed $W$ bosons in the $(\eta,\phi)$ plane,
	\begin{equation}
		\Delta R_{WW}
		=
		\sqrt{(\Delta\eta_{WW})^2+(\Delta\phi_{WW})^2},
	\end{equation}
	where $\Delta\eta_{WW}$ and $\Delta\phi_{WW}$ denote the differences in pseudorapidity and azimuthal angle of the two reconstructed $W$ bosons. The separation is mainly governed by the boost of the parent Higgs boson: a more highly boosted Higgs produces more collimated $W$ bosons, resulting in smaller values of $\Delta R_{WW}$, whereas a Higgs produced nearly at rest yields larger separations. Since the boost distributions differ between signal and background processes, $\Delta R_{WW}$ provides useful discriminating power.
	
	\item \textbf{$\beta_H$:}
	The velocity of the reconstructed Higgs candidate,
	\begin{equation}
		\beta_H
		=
		\frac{|\vec p_H|}{E_H}
		=
		\frac{|\vec p_{W_1}+\vec p_{W_2}|}
		{E_{W_1}+E_{W_2}},
	\end{equation}
	where $\vec p_H$ and $E_H$ are the momentum and energy of the reconstructed Higgs candidate. Owing to the different production kinematics of signal and background events, the $\beta_H$ distributions exhibit distinct shapes and therefore provide additional discrimination.
	
	\item \textbf{$\cos\theta_H$:}
	The cosine of the polar production angle of the reconstructed Higgs candidate in the laboratory frame,
	\begin{equation}
		\cos\theta_H
		=
		\frac{p_{z,H}}
		{|\vec p_H|},
	\end{equation}
	where $p_{z,H}$ is the longitudinal momentum component of the reconstructed Higgs candidate. The signal and background processes possess different production-angle distributions, making this observable effective for background suppression.
	
	\item \textbf{$p_{H}$:}
	The momentum of the reconstructed Higgs candidate,
	\begin{equation}
		p_{H}
		=
		|\vec p_{H}|.
	\end{equation}
	Although this observable also provides discrimination between signal and background, it is found to be strongly correlated with $\beta_H$ (see Tables~\ref{tab:cov_hzll160}--\ref{tab:cov_ww4j2n}). It is therefore not employed as an independent selection variable in order to avoid redundant cuts.
\end{itemize}

\begin{figure}[h!]
	\centering
	\includegraphics[width=0.45\textwidth]{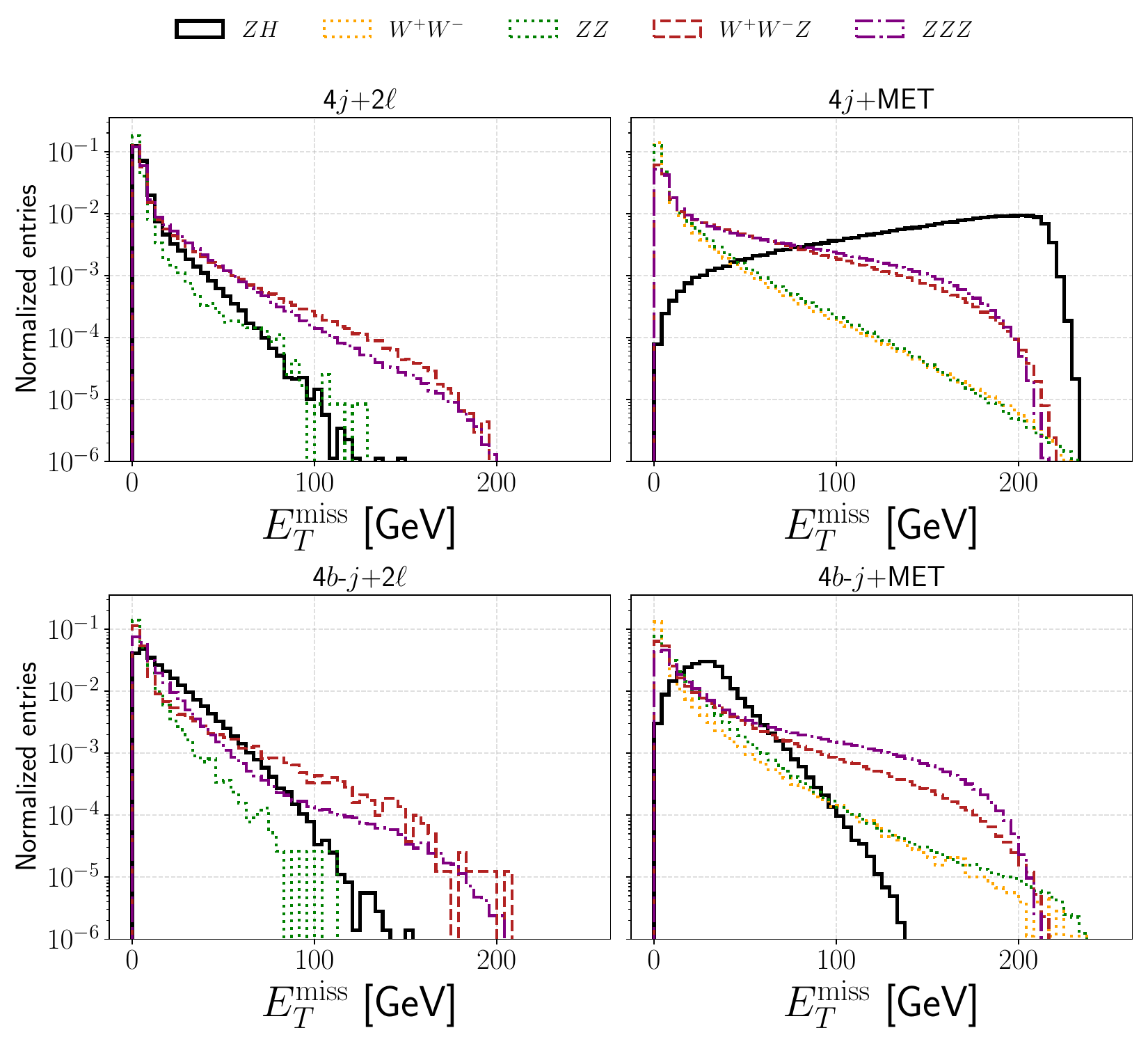}
	\caption{Missing transverse energy distribution for signal and background processes. The Higgs boson masses $m_H=160$ and $400$ GeV are chosen for the upper and lower rows.}
	\label{fig:met}
\end{figure}

\begin{figure}[h!]
	\centering
	\includegraphics[width=0.45\textwidth]{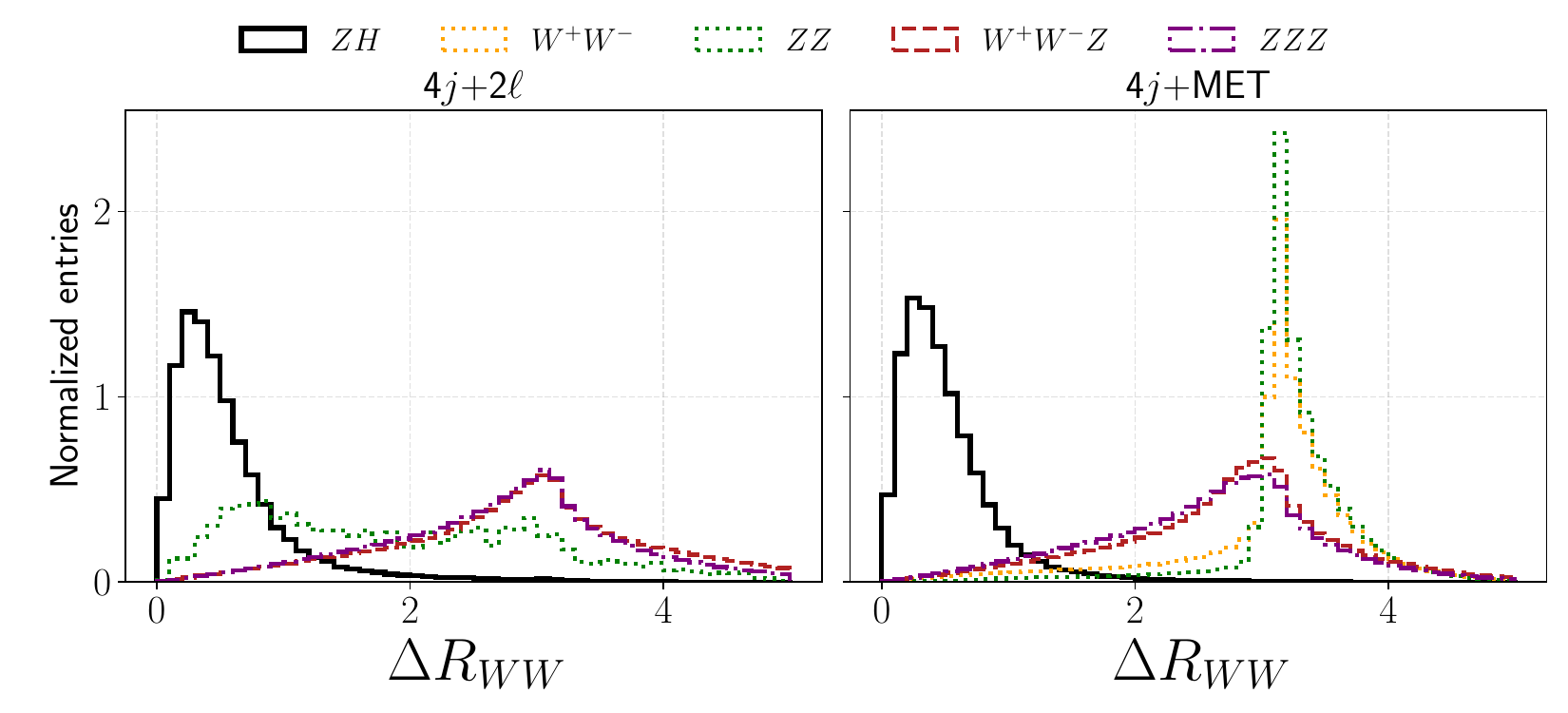}
	\caption{$\Delta R_{WW}$ distribution for signal and background processes. The Higgs boson mass is set to $m_H=160$ GeV.}
	\label{fig:DR}
\end{figure}

\begin{figure}[h!]
	\centering
	\includegraphics[width=0.45\textwidth]{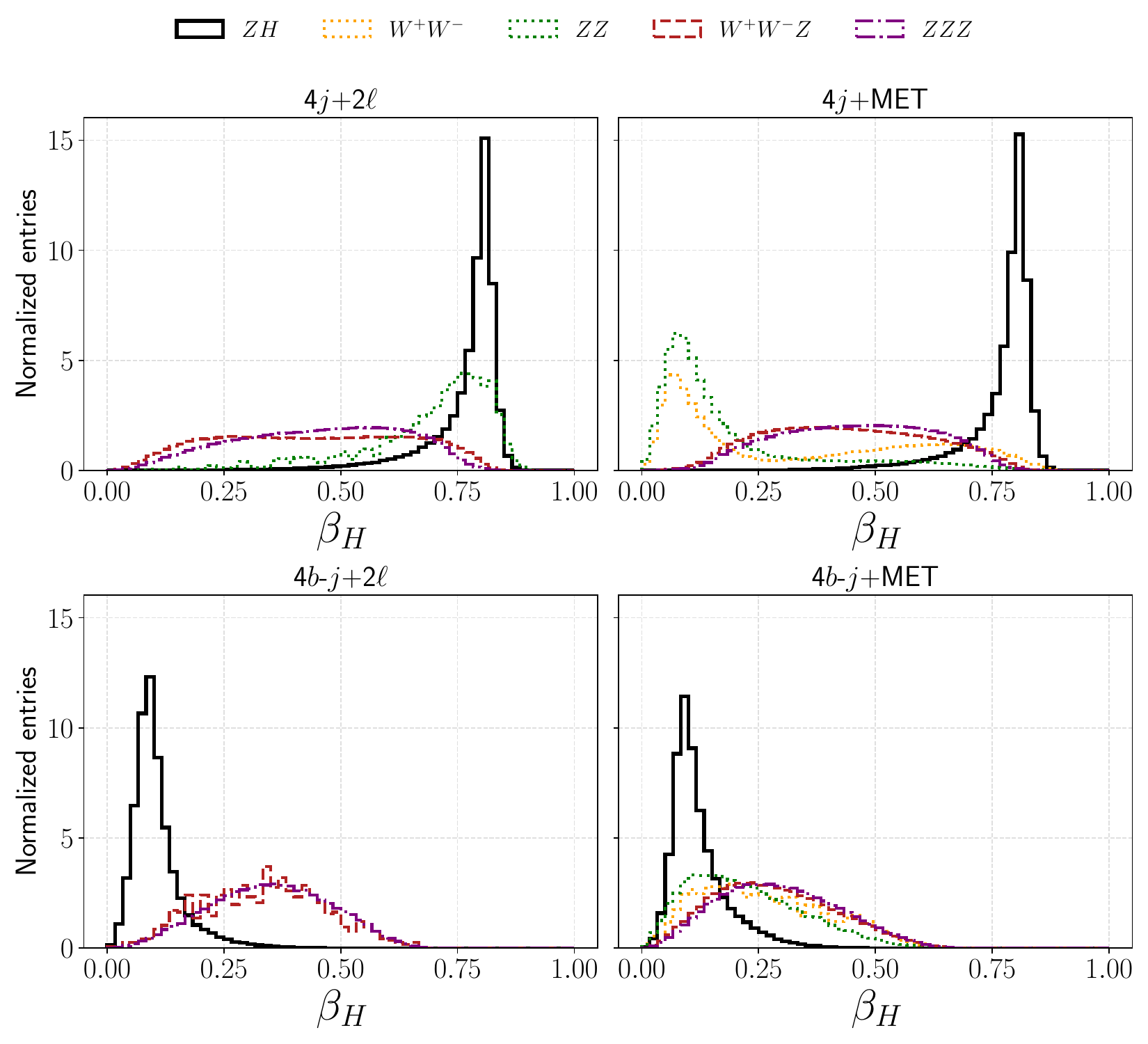}
	\caption{$\beta_H$ distribution for signal and background processes. The Higgs boson masses $m_H=160$ and $400$ GeV are chosen for the upper and lower rows.}
	\label{fig:betaH}
\end{figure}

\begin{figure}[h!]
	\centering
	\includegraphics[width=0.45\textwidth]{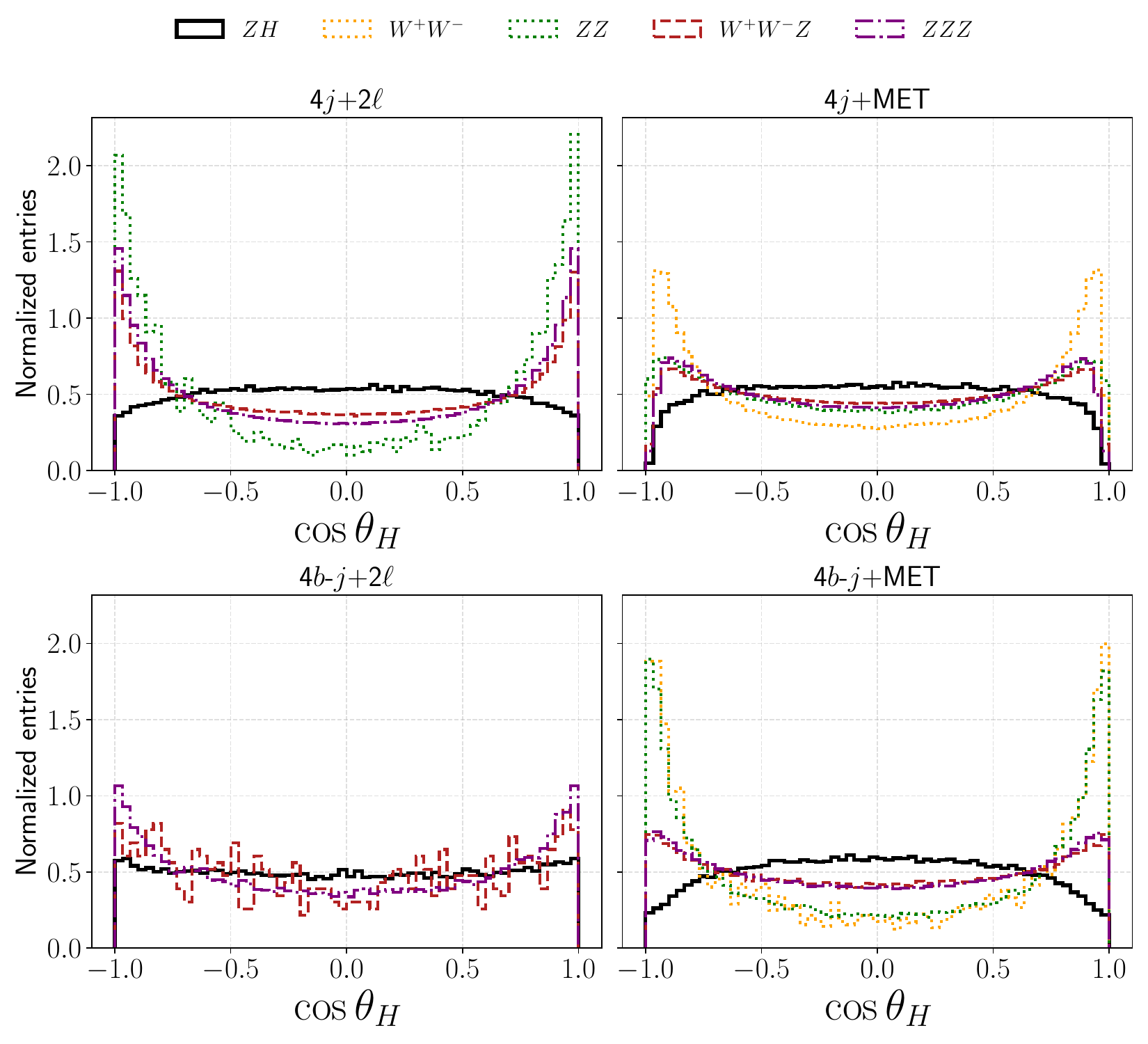}
	\caption{$\cos\theta_H$ distribution for signal and background processes. The Higgs boson masses $m_H=160$ and $400$ GeV are chosen for the upper and lower rows.}
	\label{fig:costH}
\end{figure}

The distributions of the key discriminating variables, $E_T^{\rm miss}$, $\Delta R_{WW}$, $\beta_H$, and $\cos\theta_H$ are shown in Figures~\ref{fig:met},~\ref{fig:DR},~\ref{fig:betaH}, and~\ref{fig:costH}, illustrating the separation power between signal and the main background processes.

It is also observed that these distributions vary with the Higgs boson mass. To preserve signal statistics at high masses, particularly near the center-of-mass energy threshold, the cuts on $\Delta R_{WW}$, $\beta_H$, and $\cos\theta_H$ are kept conservative in the 4$j$ channels and are omitted entirely in the 4$b$-$j$ channels, where signal yields are already limited by the lower cross section, $b$-tagging and $h$ mass window requirements.

The correlations among the variables $\Delta R_{WW}$, $\beta_H$, $p_H$, and $\cos\theta_H$ are examined via correlation matrices for both signal and the main background processes, presented in Tables~\ref{tab:cov_hzll160}--\ref{tab:cov_ww4j2n}. The strong correlation between $\beta_H$ and $p_H$ (typically $> 0.9$) indicates that $p_H$ provides little additional discriminating power beyond $\beta_H$. Consequently, $p_H$ is omitted to avoid over-constraining the signal. 

In what follows, the event selection for each final state is described in
detail. The selection criteria are motivated by the characteristic kinematic
properties of the signal and the detector resolutions. Candidate thresholds
were determined through an iterative study of the kinematic distributions,
guided by the signal--background separation and refined through several
successive analysis iterations. The distributions presented in
\Cref{fig:met,fig:DR,fig:betaH,fig:costH} motivate the adopted selection
requirements, whose thresholds provide an effective balance between signal
efficiency and background rejection.

\begin{table}[htbp]
	\centering
	\caption{Correlation matrix for signal (4$j$+2$\ell$)}
	\begin{tabular}{l|cccc}
		\hline
		& $\Delta R_{WW}$ & $\beta_H$ & $p_H$ & $\cos\theta_H$ \\
		\hline
		$\Delta R_{WW}$ & 1.00 & \num{-4.79e-1} & \num{-2.90e-1} & \num{-1.22e-3} \\
		$\beta_H$      & \num{-4.79e-1} & 1.00 & \num{9.17e-1} & \num{3.96e-3} \\
		$p_H$          & \num{-2.90e-1} & \num{9.17e-1} & 1.00 & \num{3.42e-3} \\
		$\cos\theta_H$ & \num{-1.22e-3} & \num{3.96e-3} & \num{3.42e-3} & 1.00 \\
		\hline
	\end{tabular}
	\label{tab:cov_hzll160}
\end{table}
\begin{table}[htbp]
	\centering
	\caption{Correlation matrix for signal (4$j$+$E_T^{miss}$)}
	\begin{tabular}{l|cccc}
		\hline
		& $\Delta R_{WW}$ & $\beta_H$ & $p_H$ & $\cos\theta_H$ \\
		\hline
		$\Delta R_{WW}$ & 1.00 & \num{-4.45e-1} & \num{-2.22e-1} & \num{-3.64e-3} \\
		$\beta_H$      & \num{-4.45e-1} & 1.00 & \num{9.18e-1} & \num{-4.20e-4} \\
		$p_H$          & \num{-2.22e-1} & \num{9.18e-1} & 1.00 & \num{-1.56e-3} \\
		$\cos\theta_H$ & \num{-3.64e-3} & \num{-4.20e-4} & \num{-1.56e-3} & 1.00 \\
		\hline
	\end{tabular}
	\label{tab:cov_hznn160}
\end{table}
\begin{table}[htbp]
	\centering
	\caption{Correlation matrix for $WWZ$ (4$j$+2$\ell$)}
	\begin{tabular}{l|cccc}
		\hline
		& $\Delta R_{WW}$ & $\beta_H$ & $p_H$ & $\cos\theta_H$ \\
		\hline
		$\Delta R_{WW}$ & 1.00 & \num{-7.46e-1} & \num{-6.58e-1} & \num{1.25e-4} \\
		$\beta_H$      & \num{-7.46e-1} & 1.00 & \num{9.41e-1} & \num{-5.90e-4} \\
		$p_H$          & \num{-6.58e-1} & \num{9.41e-1} & 1.00 & \num{-3.81e-4} \\
		$\cos\theta_H$ & \num{1.25e-4} & \num{-5.90e-4} & \num{-3.81e-4} & 1.00 \\
		\hline
	\end{tabular}
	\label{tab:cov_wwz4j2l}
\end{table}
\begin{table}[htbp]
	\centering
	\caption{Correlation matrix for WW (4$j$+$E_T^{miss}$)}
	\begin{tabular}{l|cccc}
		\hline
		& $\Delta R_{WW}$ & $\beta_H$ & $p_H$ & $\cos\theta_H$ \\
		\hline
		$\Delta R_{WW}$ & 1.00 & \num{-4.98e-1} & \num{-4.04e-1} & \num{-2.15e-3} \\
		$\beta_H$      & \num{-4.98e-1} & 1.00 & \num{9.63e-1} & \num{2.00e-3} \\
		$p_H$          & \num{-4.04e-1} & \num{9.63e-1} & 1.00 & \num{1.33e-3} \\
		$\cos\theta_H$ & \num{-2.15e-3} & \num{2.00e-3} & \num{1.33e-3} & 1.00 \\
		\hline
	\end{tabular}
	\label{tab:cov_ww4j2n}
\end{table}

\subsection{$Z \to \ell^+\ell^-,\; H \to WW \to 4j$ channel (4j+2$\ell$)}

Events are required to contain at least four reconstructed light jets with a light-jet veto applied to suppress $b$-jet contributions from $t\bar{t}$ and $t\bar{t}b\bar{b}$ processes. At least two isolated leptons must satisfy
\begin{equation}
	p_T > 10~\mathrm{GeV}, \qquad |\eta| < 5 .
\end{equation}

The dilepton invariant mass must be consistent with an on-shell $Z$ boson,
\begin{equation}
	|m_{\ell\ell} - m_Z| < 10~\mathrm{GeV}.
\end{equation}
A cut on the missing transverse energy is applied to suppress backgrounds with genuine missing energy,
\begin{equation}
	E_T^{\rm miss} < 20~\mathrm{GeV}.
\end{equation}

The two $W$ bosons are reconstructed from the four highest-$E_T$ light jets in the event. For each pairing of jets, the dijet invariant masses $m_{j_ij_j}$ and $m_{j_kj_l}$ are computed, and the combination minimizing
\begin{equation}
	|m_{j_ij_j} - m_W|+|m_{j_kj_l} - m_W|
\end{equation}
is selected. Both candidates are then required to lie within a mass window around the nominal $W$ mass,
\begin{equation}
	|m_{jj} - m_W| < 30~\mathrm{GeV}.
\end{equation}

A simple mass-constrained correction is applied to the jets forming each reconstructed $W$ candidate by scaling their four-momenta according to
\begin{equation}
	p_i^{\mathrm{corr}}
	=
	\frac{m_W^{\mathrm{PDG}}}{m_{jj}}
	p_i,
	\qquad i=1,2,
\end{equation}
where $m_{jj}$ is the reconstructed invariant mass of the dijet system, $p_i$ is the four-momentum of the $i$-th jet, and $m_W^{\mathrm{PDG}}=80.38~\mathrm{GeV}$ is the nominal $W$-boson mass. Both jet four-momenta are rescaled by the same factor, thereby preserving their directions while constraining the corrected dijet invariant mass to $m_W^{\mathrm{PDG}}$. The corrected $W$-boson four-momentum is then reconstructed as the sum of the corrected jet four-momenta.

The Higgs candidate is then reconstructed from the four-vector sum of the two corrected $W$ bosons,
\begin{equation}
	p_H = p_{W_1}^{\mathrm{corr}} + p_{W_2}^{\mathrm{corr}}.
\end{equation}

To further suppress backgrounds, cuts are applied on three kinematic variables derived from the $W$ boson four-momenta:
\begin{equation}
	\Delta R_{WW} < 2.0, \qquad \beta_H > 0.4, \qquad \cos\theta_H < 0.75,
\end{equation}
These cuts are motivated by the correlation matrices and the distributions shown in Figures~\ref{fig:DR},~\ref{fig:betaH}, and~\ref{fig:costH}, which demonstrate good separation between signal and background. The Higgs candidate mass is finally calculated as
\begin{equation}
	m_H = m(W_1^{\mathrm{corr}} + W_2^{\mathrm{corr}}).
\end{equation}

\subsection{$Z \to \nu\bar{\nu},\; H \to WW \to 4j$ channel (4j+$E_T^{miss}$)}

This channel follows the same jet selection and $W$ reconstruction procedure as the 4j+2$\ell$ channel. A lepton veto is imposed to reject events with charged leptons, and the missing transverse momentum must satisfy
\begin{equation}
	E_T^{\rm miss} > 50~\mathrm{GeV}.
\end{equation}

The two $W$ bosons are reconstructed from the four highest-$E_T$ light jets using the same pairing algorithm and $W$ mass window as in the 4j+2$\ell$ channel,
\begin{equation}
	|m_{jj} - m_W| < 30~\mathrm{GeV}.
\end{equation}

The same kinematic cuts are applied to suppress backgrounds:
\begin{equation}
	\Delta R_{WW} < 2.0, \qquad \beta_H > 0.4, \qquad \cos\theta_H < 0.75.
\end{equation}
The Higgs candidate is reconstructed as described above from the four-vector sum of the two corrected $W$ bosons.
\subsection{$Z \to \ell^+\ell^-,\; H \to hh \to 4b$ channel (4bj+2$\ell$)}

Events are required to contain at least four $b$-tagged jets (90\% tagging efficiency) and two isolated leptons with
\begin{equation}
	p_T > 10~\mathrm{GeV}, \qquad |\eta| < 5 .
\end{equation}

The dilepton invariant mass must satisfy the $Z$ mass window,
\begin{equation}
	|m_{\ell\ell} - m_Z| < 10~\mathrm{GeV},
\end{equation}
and a cut on the missing transverse energy is applied,
\begin{equation}
	E_T^{\rm miss} < 20~\mathrm{GeV}.
\end{equation}

Two light Higgs candidates are reconstructed from all distinct $b$-jet pairings by minimizing
\begin{equation}
	|m_{b_ib_j}-m_h| + |m_{b_kb_l}-m_h|.
\end{equation}
Both candidates are required to lie within a mass window around the nominal $h$ boson mass,
\begin{equation}
	|m_{bb} - m_h| < 30~\mathrm{GeV}.
\end{equation}

A similar mass-constrained correction is applied to the $b$-jets forming each reconstructed $h$ candidate. In this case, the jet four-momenta are rescaled using the nominal Higgs boson mass,
\begin{equation}
	p_i^{\mathrm{corr}}
	=
	\frac{m_h^{\mathrm{PDG}}}{m_{bb}}
	p_i,
\end{equation}
where $m_{bb}$ is the reconstructed invariant mass of the corresponding $b$-jet pair and $m_h^{\mathrm{PDG}}=125~\mathrm{GeV}$ is the nominal Higgs boson mass. The corrected Higgs candidates are then reconstructed from the sums of the corrected $b$-jet four-momenta, and the heavy Higgs candidate is obtained as
\begin{equation}
	p_H
	=
	p_{h_1}^{\mathrm{corr}}
	+
	p_{h_2}^{\mathrm{corr}}.
\end{equation}

The heavy Higgs candidate mass is then reconstructed from the sum of the two light Higgs four-momenta,
\begin{equation}
	m_H = m(h_1^{\mathrm{corr}} + h_2^{\mathrm{corr}}).
\end{equation}
No cuts on $\Delta R_{WW}$, $\beta_H$, or $\cos\theta_H$ are applied in this channel to preserve signal statistics.

\subsection{$Z \to \nu\bar{\nu},\; H \to hh \to 4b$ channel (4bj+$E_T^{miss}$)}

This channel follows the same $b$-jet selection as the 4bj+2$\ell$ channel. At least four $b$-tagged jets (90\% tagging efficiency) are required, a lepton veto is imposed, and events must satisfy
\begin{equation}
	E_T^{\rm miss} > 20~\mathrm{GeV}.
\end{equation}

Two light Higgs candidates are reconstructed from all distinct $b$-jet pairings by minimizing
\begin{equation}
	|m_{b_ib_j}-m_h| + |m_{b_kb_l}-m_h|,
\end{equation}
with both candidates required to satisfy
\begin{equation}
	|m_{bb} - m_h| < 30~\mathrm{GeV}.
\end{equation}
The Higgs candidate is reconstructed as described above from the four-vector sum of the two corrected $h$ bosons.

No cuts on $\Delta R_{WW}$, $\beta_H$, or $\cos\theta_H$ are applied in this channel either to preserve signal statistics.

\subsection{Cut efficiencies}

For all channels, selection efficiencies are computed after each selection step, allowing a transparent assessment of signal acceptance and background rejection at each stage of the analysis. The selection sequences for the four final states are detailed in the previous section, with the resulting efficiencies together with event yields at $\cba=0.1$ and $\mathcal{L}=10~ \ab$ are presented in Tables~\ref{tab:signal_4j2l_light}--\ref{tab:bkg_4bj0l}.

The jet multiplicity requirements provide the first level of background suppression, effectively rejecting QCD multijet events with low jet multiplicity. In the 4j channels, the light-jet veto significantly reduces contributions from $t\bar{t}$ and $t\bar{t}b\bar{b}$, where the presence of $b$-jets leads to rejection. The dilepton mass window and the $E_T^{miss}$ cuts are particularly powerful against $WW$ and $ZZ$ backgrounds, which either lack leptons or have different missing energy characteristics.

The $W$ mass window in the 4j channels further suppresses non-resonant backgrounds, while the subsequent cuts on $\Delta R_{WW}$, $\beta_H$, and $\cos\theta_H$ provide the final and most powerful background rejection. These kinematic variables exploit the spin correlations and boost structure of the $H \to WW$ decay, which differ significantly between signal and background processes. As demonstrated by the correlation matrices in Tables~\ref{tab:cov_hzll160}--\ref{tab:cov_ww4j2n}, the combination of these variables efficiently separates the signal from $WWZ$ and $ZZZ$ backgrounds.

In the 4bj channels, the $b$-tagging requirement provides excellent suppression of light-jet backgrounds, while the $h$ mass window effectively isolates the $h \to b\bar{b}$ decays. The cuts on $\Delta R_{WW}$, $\beta_H$, and $\cos\theta_H$ are not applied in these channels to preserve signal statistics, as the $b$-tagging already provides sufficient background rejection.
\begin{table*}[htbp]
	\centering
	\caption{Signal selection for 4$j$+2$\ell$ final state ($\cba=0.1$) -- efficiency (remaining events at $10~ \ab$)}
	\begin{tabular}{l|c|c|c|c|c}
		\hline
		$m_H$ [GeV] & 160 & 180 & 200 & 220 & 240 \\
		\hline
		Original $\sigma$ (fb) & 0.013 & 0.012 & 0.0088 & 0.0076 & 0.0067 \\
		\hline
		Cuts & \multicolumn{5}{c}{Selection efficiencies (remaining events)}\\
		\hline
		4 jets    & 0.50 (64) & 0.61 (75) & 0.71 (62) & 0.76 (58) & 0.79 (53) \\
		2 Leptons & 0.62 (40) & 0.64 (48) & 0.64 (40) & 0.63 (37) & 0.62 (33) \\
		$Z$ cut   & 0.99 (39) & 0.99 (47) & 0.99 (40) & 0.99 (36) & 0.99 (33) \\
		$E_T^{miss}$ & 0.94 (37) & 0.93 (44) & 0.92 (37) & 0.92 (33) & 0.91 (30) \\
		$W$ cut   & 0.60 (22) & 0.71 (31) & 0.81 (30) & 0.86 (29) & 0.89 (27) \\
		$\Delta R_{WW}$ & 0.97 (22) & 0.93 (29) & 0.86 (26) & 0.70 (20) & 0.50 (13) \\
		$\beta_H$ & 0.99 (22) & 1.00 (29) & 1.00 (26) & 1.00 (20) & 1.00 (13) \\
		$\cos\theta_H$ & 0.80 (17) & 0.85 (25) & 0.90 (23) & 0.94 (19) & 0.96 (13) \\
		\hline
		Total & 0.14 (17) & 0.20 (25) & 0.26 (23) & 0.25 (19) & 0.19 (13) \\
		\hline
	\end{tabular}
	\label{tab:signal_4j2l_light}
\end{table*}
\begin{table*}[htbp]
	\centering
	\caption{Signal selection for 4$j$+$E_T^{miss}$ final state ($\cba=0.1$) -- efficiency (remaining events at $10~ \ab$)}
	\begin{tabular}{l|c|c|c|c|c}
		\hline
		$m_H$ [GeV] & 160 & 180 & 200 & 220 & 240 \\
		\hline
		Original $\sigma$ (fb) & 0.042 & 0.041 & 0.029 & 0.026 & 0.022 \\
		\hline
Cuts & \multicolumn{5}{c}{Selection efficiencies (remaining events)}\\
\hline
		4 jets    & 0.42 (180) & 0.54 (220) & 0.64 (190) & 0.71 (180) & 0.75 (170) \\
		Lepton veto & 1.00 (180) & 1.00 (220) & 1.00 (190) & 1.00 (180) & 1.00 (170) \\
		$E_T^{miss}$ & 0.95 (170) & 0.96 (210) & 0.96 (180) & 0.96 (170) & 0.95 (160) \\
		$W$ cut   & 0.58 (100) & 0.70 (150) & 0.79 (140) & 0.85 (150) & 0.88 (140) \\
		$\Delta R_{WW}$ & 0.99 (99) & 0.96 (140) & 0.88 (130) & 0.72 (110) & 0.51 (72) \\
		$\beta_H$ & 1.00 (99) & 1.00 (140) & 1.00 (130) & 1.00 (110) & 1.00 (72) \\
		$\cos\theta_H$ & 0.82 (81) & 0.85 (120) & 0.90 (110) & 0.94 (100) & 0.96 (69) \\
		\hline
		Total & 0.19 (81) & 0.29 (120) & 0.39 (110) & 0.39 (100) & 0.31 (69) \\
		\hline
	\end{tabular}
	\label{tab:signal_4j0l_light}
\end{table*}
\begin{table*}[htbp]
	\centering
	\caption{Signal selection for 4$b$-$j$+2$\ell$ final state ($\cba=0.1$) -- efficiency (remaining events at $10~ \ab$)}
	\begin{tabular}{l|c|c|c|c|c|c|c|c}
		\hline
		$m_H$ [GeV] & 260 & 280 & 300 & 320 & 340 & 360 & 380 & 400 \\
		\hline
		Original $\sigma$ (fb) & \num{2.4e-3} & \num{2.7e-3} & \num{2.5e-3} & \num{2.1e-3} & \num{1.7e-3} & \num{1.2e-3} & \num{8.0e-4} & \num{4.0e-4} \\
				\hline
		Cuts & \multicolumn{8}{c}{Selection efficiencies (remaining events)}\\
		\hline
		4 bjets   & 0.72 (17) & 0.75 (20) & 0.77 (19) & 0.79 (17) & 0.80 (14) & 0.81 (10) & 0.82 (6.5) & 0.82 (3.3) \\
		2 Leptons & 0.62 (11) & 0.59 (12) & 0.56 (11) & 0.53 (8.8) & 0.51 (6.9) & 0.48 (4.9) & 0.46 (3.0) & 0.44 (1.5) \\
		$Z$ cut   & 0.99 (11) & 0.99 (12) & 0.99 (11) & 0.99 (8.7) & 0.99 (6.8) & 0.99 (4.8) & 0.99 (3.0) & 0.99 (1.4) \\
		$E_T^{miss}$ & 0.79 (8.4) & 0.78 (9.2) & 0.77 (8.2) & 0.75 (6.5) & 0.74 (5.0) & 0.72 (3.5) & 0.71 (2.1) & 0.70 (1.0) \\
		$m_h$ cut & 0.32 (2.7) & 0.34 (3.1) & 0.39 (3.2) & 0.43 (2.8) & 0.47 (2.4) & 0.50 (1.7) & 0.51 (1.1) & 0.52 (0.52) \\
		\hline
		Total & 0.11 (2.7) & 0.12 (3.1) & 0.13 (3.2) & 0.13 (2.8) & 0.14 (2.4) & 0.14 (1.7) & 0.13 (1.1) & 0.13 (0.52) \\
		\hline
	\end{tabular}
	\label{tab:signal_4bj2l_heavy}
\end{table*}
\begin{table*}[htbp]
	\centering
	\caption{Signal selection for 4$b$-$j$+$E_T^{miss}$ final state ($\cba=0.1$) -- efficiency (remaining events at $10~ \ab$)}
	\begin{tabular}{l|c|c|c|c|c|c|c|c}
		\hline
		$m_H$ [GeV] & 260 & 280 & 300 & 320 & 340 & 360 & 380 & 400 \\
		\hline
		Original $\sigma$ (fb) & \num{8.1e-3} & \num{9.0e-3} & \num{8.2e-3} & \num{7.0e-3} & \num{5.6e-3} & \num{4.2e-3} & \num{2.7e-3} & \num{1.3e-3} \\
		\hline
Cuts & \multicolumn{8}{c}{Selection efficiencies (remaining events)}\\
\hline
		4 bjets   & 0.70 (57) & 0.73 (66) & 0.75 (62) & 0.77 (54) & 0.78 (44) & 0.79 (33) & 0.80 (22) & 0.81 (11) \\
		Lepton veto & 1.00 (57) & 1.00 (66) & 1.00 (62) & 1.00 (54) & 1.00 (44) & 1.00 (33) & 1.00 (22) & 1.00 (11) \\
		$E_T^{miss}$ & 0.99 (56) & 0.99 (65) & 0.99 (61) & 0.98 (53) & 0.97 (43) & 0.95 (31) & 0.91 (20) & 0.73 (7.9) \\
		$m_h$ cut & 0.28 (16) & 0.30 (20) & 0.34 (21) & 0.38 (20) & 0.42 (18) & 0.45 (14) & 0.47 (9.4) & 0.47 (3.7) \\
		\hline
		Total & 0.19 (16) & 0.22 (20) & 0.25 (21) & 0.29 (20) & 0.32 (18) & 0.34 (14) & 0.34 (9.4) & 0.27 (3.7) \\
		\hline
	\end{tabular}
	\label{tab:signal_4bj0l_heavy}
\end{table*}
\begin{table*}[htbp]
	\centering
	\caption{Background selection for 4$j$+2$\ell$ final state -- efficiency (remaining events at $10~ \ab$)}
	\begin{tabular}{l|c|c|c|c|c|c}
		\hline
		Process & $t\bar{t}$ & $t\bar{t}b\bar{b}$ & WW & ZZ & WWZ & ZZZ \\
		\hline
		Original $\sigma$ (fb) & 26 & \num{4.1e-2} & \num{7.9e3} & 460 & 34 & 1.1 \\
		\hline
Cuts & \multicolumn{6}{c}{Selection efficiencies (remaining events)}\\
\hline
		4 jets    & 0.01 (\num{1.3e5}) & 0.01 (6.1) & 0.34 (\num{2.7e7}) & 0.27 (\num{1.2e6}) & 0.61 (\num{2.1e5}) & 0.52 (\num{5.5e3}) \\
		2 Leptons & 0.16 (\num{2.1e4}) & 0.19 (1.1) & 0.00 (160) & 0.00 (\num{2.0e3}) & 0.02 (\num{5.1e3}) & 0.06 (340) \\
		$Z$ cut   & 0.15 (\num{3.2e3}) & 0.17 (0.19) & 0.05 (8.1) & 0.95 (\num{1.9e3}) & 0.92 (\num{4.7e3}) & 0.95 (320) \\
		$E_T^{miss}$ & 0.08 (250) & 0.09 (0.016) & 0.00 (0) & 0.97 (\num{1.9e3}) & 0.88 (\num{4.2e3}) & 0.88 (280) \\
		$W$ cut   & 0.23 (57) & 0.29 (0.0048) & --- & 0.10 (180) & 0.80 (\num{3.3e3}) & 0.82 (230) \\
		$\Delta R_{WW}$ & 0.26 (15) & 0.28 (0.0014) & --- & 0.55 (97) & 0.19 (630) & 0.21 (49) \\
		$\beta_H$ & 0.86 (13) & 0.85 (0.0012) & --- & 1.00 (97) & 0.99 (630) & 0.99 (48) \\
		$\cos\theta_H$ & 0.78 (9.8) & 0.80 (\num{9.3e-4}) & --- & 0.47 (45) & 0.70 (440) & 0.72 (35) \\
		\hline
		Total & \num{3.8e-7} (9.8) & \num{2.3e-6} (\num{9.3e-4}) & 0 (0) & \num{9.9e-6} (45) & 0.0013 (440) & 0.0035 (35) \\
		\hline
	\end{tabular}
	\label{tab:bkg_4j2l}
\end{table*}
\begin{table*}[htbp]
	\centering
	\caption{Background selection for 4$j$+$E_T^{miss}$ final state -- efficiency (remaining events at $10~ \ab$)}
	\begin{tabular}{l|c|c|c|c|c|c}
		\hline
		Process & $t\bar{t}$ & $t\bar{t}b\bar{b}$ & WW & ZZ & WWZ & ZZZ \\
		\hline
		Original $\sigma$ (fb) & 26 & \num{4.1e-2} & \num{7.9e3} & 460 & 34 & 1.1 \\
		\hline
Cuts & \multicolumn{6}{c}{Selection efficiencies (remaining events)}\\
\hline
		4 jets    & 0.01 (\num{1.3e5}) & 0.01 (6.1) & 0.34 (\num{2.7e7}) & 0.27 (\num{1.2e6}) & 0.61 (\num{2.1e5}) & 0.52 (\num{5.5e3}) \\
		Lepton veto & 0.32 (\num{4.3e4}) & 0.30 (1.8) & 1.00 (\num{2.7e7}) & 0.99 (\num{1.2e6}) & 0.76 (\num{1.6e5}) & 0.87 (\num{4.8e3}) \\
		$E_T^{miss}$ & 0.68 (\num{2.9e4}) & 0.68 (1.2) & 0.03 (\num{8.4e5}) & 0.04 (\num{4.6e4}) & 0.23 (\num{3.7e4}) & 0.27 (\num{1.3e3}) \\
		$W$ cut   & 0.44 (\num{1.3e4}) & 0.47 (0.57) & 0.44 (\num{3.7e5}) & 0.53 (\num{2.4e4}) & 0.70 (\num{2.6e4}) & 0.79 (\num{1.0e3}) \\
		$\Delta R_{WW}$ & 0.28 (\num{3.7e3}) & 0.35 (0.20) & 0.10 (\num{3.5e4}) & 0.04 (920) & 0.20 (\num{5.2e3}) & 0.25 (250) \\
		$\beta_H$ & 0.93 (\num{3.4e3}) & 0.91 (0.18) & 0.99 (\num{3.5e4}) & 0.96 (890) & 0.99 (\num{5.1e3}) & 0.99 (240) \\
		$\cos\theta_H$ & 0.89 (\num{3.0e3}) & 0.91 (0.17) & 0.28 (\num{9.9e3}) & 0.60 (530) & 0.75 (\num{3.9e3}) & 0.76 (190) \\
		\hline
		Total & \num{1.2e-4} (3000) & \num{4.0e-4} (0.17) & \num{1.3e-4} (9900) & \num{1.2e-4} (530) & 0.011 (3900) & 0.018 (190) \\
		\hline
	\end{tabular}
	\label{tab:bkg_4j0l}
\end{table*}
\begin{table*}[htbp]
	\centering
	\caption{Background selection for 4$b$-$j$+2$\ell$ final state -- efficiency (remaining events at $10~ \ab$)}
	\begin{tabular}{l|c|c|c|c|c|c}
		\hline
		Process & $t\bar{t}$ & $t\bar{t}b\bar{b}$ & WW & ZZ & WWZ & ZZZ \\
		\hline
		Original $\sigma$ (fb) & 26 & \num{4.1e-2} & \num{7.9e3} & 460 & 34 & 1.1 \\
		\hline
Cuts & \multicolumn{6}{c}{Selection efficiencies (remaining events)}\\
\hline
		4 bjets   & 0.02 (\num{5.9e5}) & 0.43 (180) & 0.01 (\num{5.0e5}) & 0.05 (\num{2.2e5}) & 0.06 (\num{2.2e4}) & 0.14 (\num{1.5e3}) \\
		2 Leptons & 0.42 (\num{2.5e5}) & 0.48 (85) & 0.00 (6.0) & 0.00 (640) & 0.01 (120) & 0.04 (67) \\
		$Z$ cut   & 0.29 (\num{7.1e4}) & 0.32 (28) & 0.17 (1.0) & 0.97 (620) & 0.83 (100) & 0.96 (64) \\
		$E_T^{miss}$ & 0.07 (\num{4.9e3}) & 0.08 (2.2) & 0.00 (0) & 0.95 (590) & 0.83 (86) & 0.83 (53) \\
		$m_h$ cut & 0.01 (44) & 0.08 (0.18) & --- & 0.00 (0) & 0.04 (3.7) & 0.08 (4.4) \\
		\hline
		Total & \num{1.7e-6} (44) & \num{4.3e-4} (0.18) & 0 (0) & 0 (0) & \num{1.1e-5} (3.7) & \num{4.2e-4} (4.4) \\
		\hline
	\end{tabular}
	\label{tab:bkg_4bj2l}
\end{table*}
\begin{table*}[htbp]
	\centering
	\caption{Background selection for 4$b$-$j$+$E_T^{miss}$ final state -- efficiency (remaining events at $10~ \ab$)}
	\begin{tabular}{l|c|c|c|c|c|c}
		\hline
		Process & $t\bar{t}$ & $t\bar{t}b\bar{b}$ & WW & ZZ & WWZ & ZZZ \\
		\hline
		Original $\sigma$ (fb) & 26 & \num{4.1e-2} & \num{7.9e3} & 460 & 34 & 1.1 \\
		\hline
Cuts & \multicolumn{6}{c}{Selection efficiencies (remaining events)}\\
\hline
		4 bjets   & 0.02 (\num{5.9e5}) & 0.43 (180) & 0.01 (\num{5.0e5}) & 0.05 (\num{2.2e5}) & 0.06 (\num{2.2e4}) & 0.14 (\num{1.5e3}) \\
		Lepton veto & 0.11 (\num{6.7e4}) & 0.10 (18) & 0.99 (\num{5.0e5}) & 0.99 (\num{2.2e5}) & 0.86 (\num{1.9e4}) & 0.91 (\num{1.4e3}) \\
		$E_T^{miss}$ & 0.95 (\num{6.3e4}) & 0.94 (17) & 0.11 (\num{5.6e4}) & 0.19 (\num{4.2e4}) & 0.29 (\num{5.5e3}) & 0.38 (530) \\
		$m_h$ cut & 0.08 (\num{4.9e3}) & 0.17 (2.9) & 0.06 (\num{3.4e3}) & 0.08 (\num{3.5e3}) & 0.09 (490) & 0.12 (64) \\
		\hline
		Total & \num{1.9e-4} (4900) & \num{7.0e-3} (2.9) & \num{4.3e-5} (3400) & \num{7.6e-4} (3500) & \num{1.4e-3} (490) & \num{6.1e-3} (64) \\
		\hline
	\end{tabular}
	\label{tab:bkg_4bj0l}
\end{table*}
\section{Results}
\begin{figure}[h!]
\centering
\includegraphics[width=0.45\textwidth]{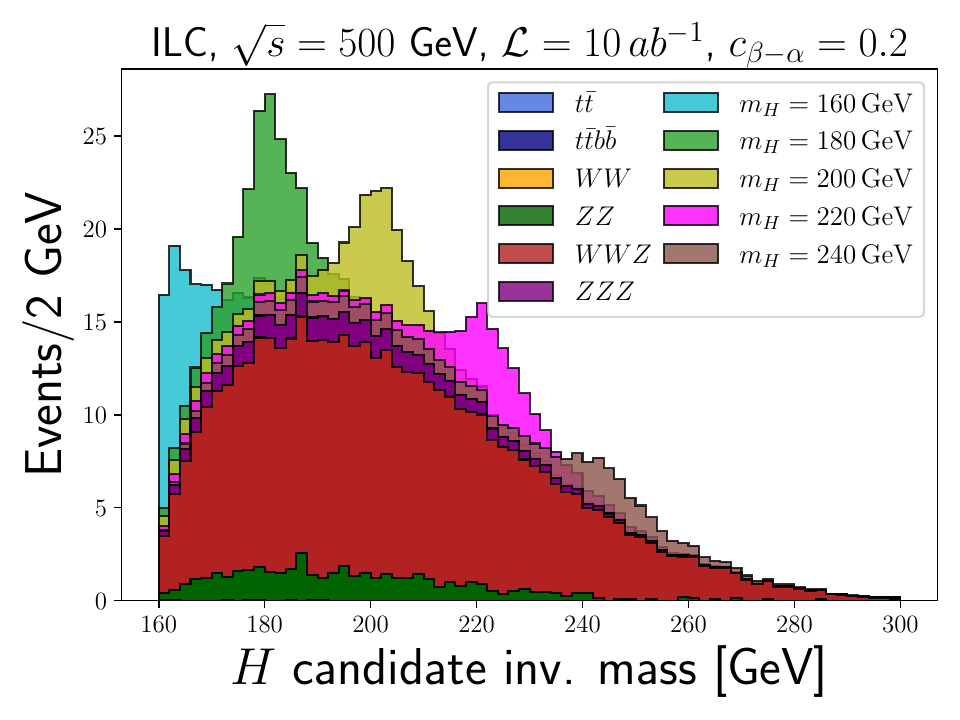}
\caption{Reconstructed invariant mass distribution for the Higgs decay channel
$H \to WW \to 4j$ with $Z\to \ell \ell$.}
\label{fig:lightll}
\end{figure}

\begin{figure}[htb!]
\centering
\includegraphics[width=0.45\textwidth]{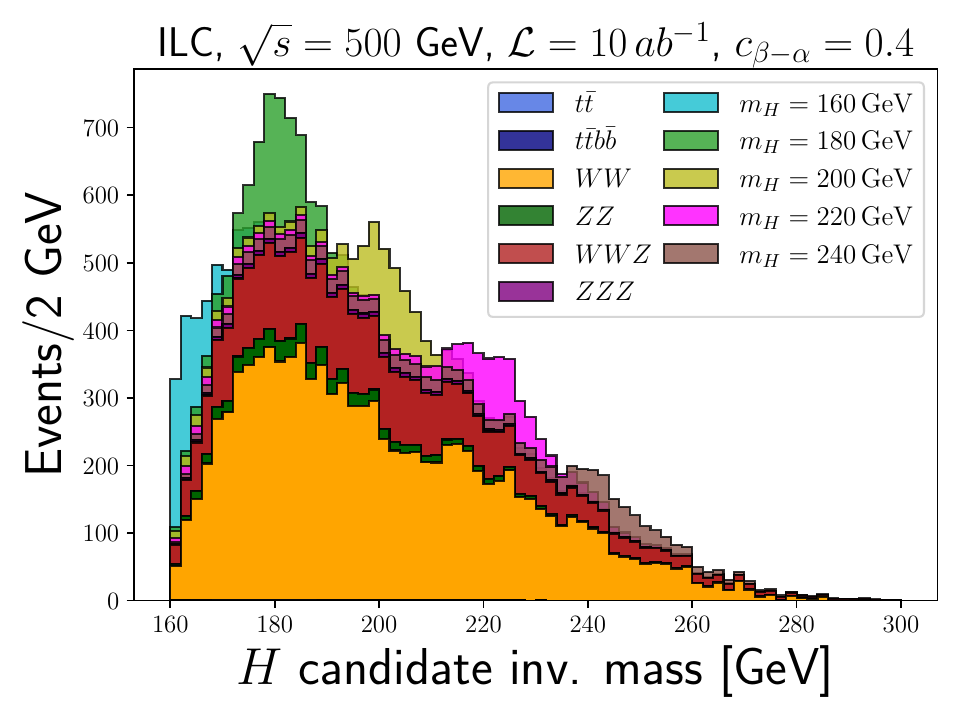}
\caption{Reconstructed invariant mass distribution for the Higgs decay channel
	$H \to WW \to 4j$ with $Z\to \nu \nu$.}
\label{fig:lightnn}
\end{figure}

\begin{figure}[h]
	\centering
	\includegraphics[width=0.45\textwidth]{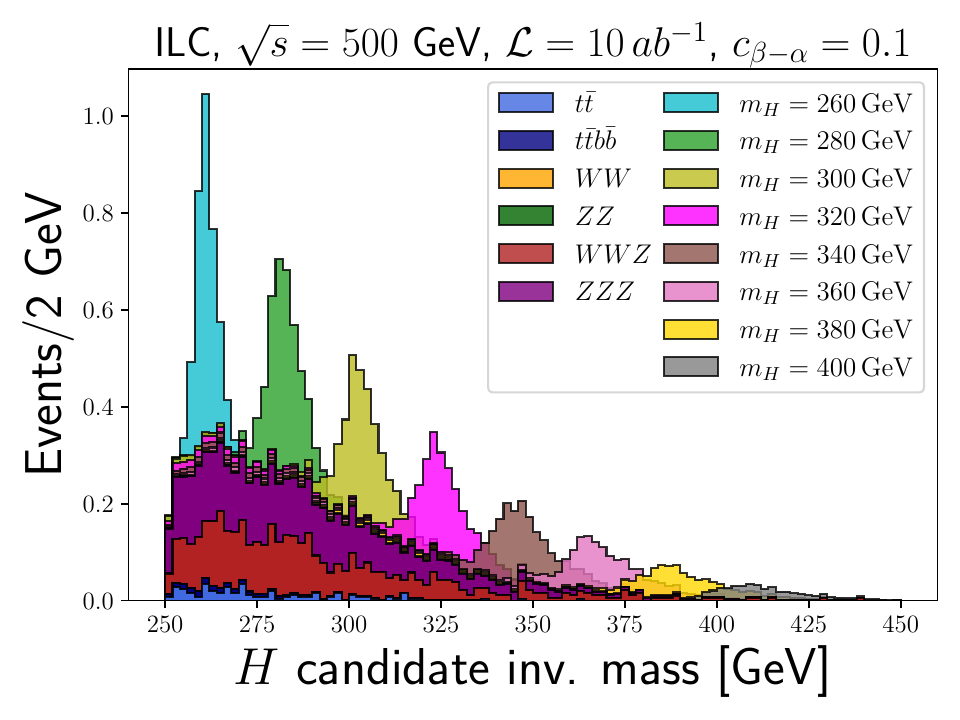}
	\caption{Reconstructed invariant mass distribution for the Higgs decay channel
		$H \to hh \to 4b$ with $Z\to \ell \ell$.}
	\label{fig:heavyll}
\end{figure}

\begin{figure}[h]
	\centering
	\includegraphics[width=0.45\textwidth]{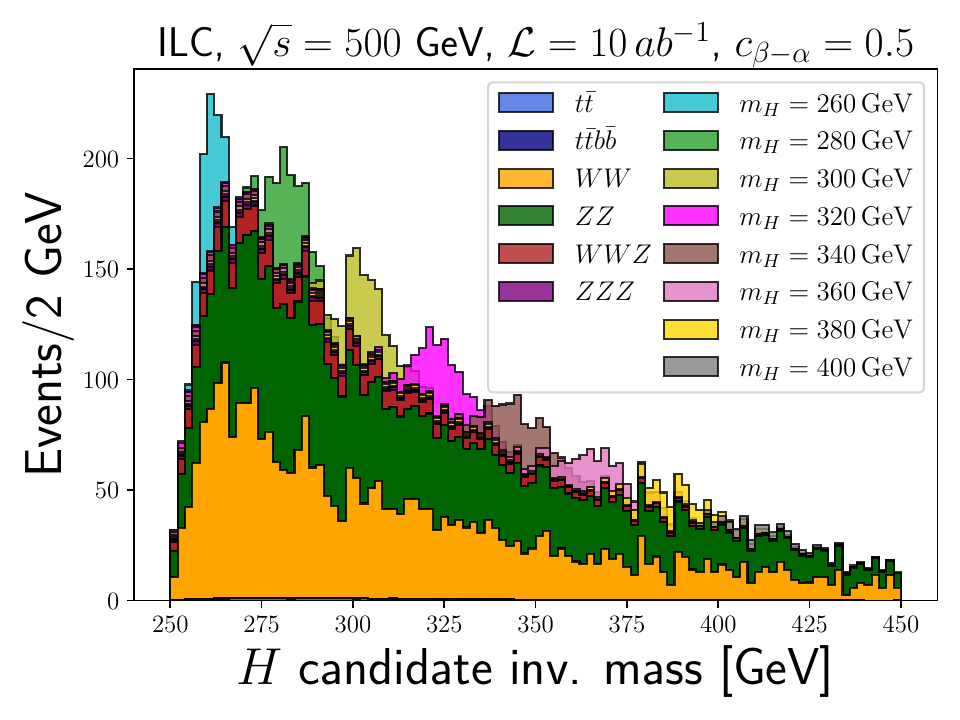}
	\caption{Reconstructed invariant mass distribution for the Higgs decay channel
		$H \to hh \to 4b$ with $Z\to \nu \nu$.}
	\label{fig:heavynn}
\end{figure}

\Cref{fig:lightll,fig:lightnn} show the reconstructed Higgs mass distributions in the 160 to 240 GeV region on top of the background processes after all selections.
The corresponding heavy Higgs mass distributions in the 260 to 400 GeV are shown in \cref{fig:heavyll,fig:heavynn}. A clear separation between signal and background is observed for both light and heavy Higgs scenarios in the di-lepton channel. However, in the final states containing missing energy, there is large background from $WW$, $ZZ$ and $ZZZ$ which are dominant in the signal region. Therefore different $\cba$ values are used to show the signal on top of the background in the four figures. The parameter choice for $\tb$ is 5 in all figures.

The exclusion contours presented in the following section are derived from
the expected statistical significance after all selection requirements. For
each benchmark mass, the signal significance obtained from the collider
simulation is taken as a reference value and subsequently rescaled across the
$(c_{\beta-\alpha},\tb)$ parameter plane according to the corresponding
production cross section and Higgs branching ratio predicted by the Type-I
2HDM. The background event yield is assumed to remain unchanged, while the
signal yield is scaled as
\begin{equation}
	S(c_{\beta-\alpha},\tb)
	=
	S_{\rm ref}
	\left(
	\frac{\sigma(c_{\beta-\alpha},\tb)}
	{\sigma_{\rm ref}}
	\right)
	\left(
	\frac{\mathrm{BR}(c_{\beta-\alpha},\tb)}
	{\mathrm{BR}_{\rm ref}}
	\right),
\end{equation}
where $\sigma_{\rm ref}$ and $\mathrm{BR}_{\rm ref}$ denote the production
cross section and branching ratio at the reference benchmark point. The
expected significance is then evaluated using the Asimov significance \cite{Cowan:2010js}:
\begin{equation}
	Z_A=
	\sqrt{
		2\left[
		(S+B)\ln\left(1+\frac{S}{B}\right)-S
		\right]
	},
\end{equation}
where $B$ denotes the expected background yield after the full event
selection. Parameter points satisfying $Z_A\ge1.64$ are taken to define the
expected one-sided $95\%$ confidence level exclusion contours.

\begin{figure}[thb]
	\centering
	\includegraphics[width=0.45\textwidth]{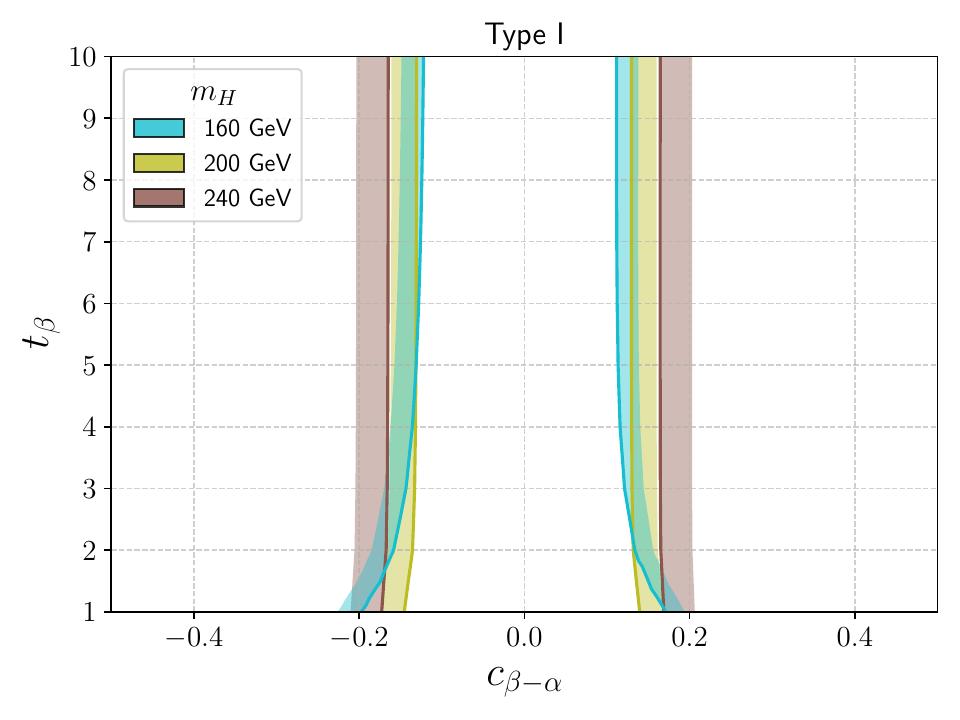}
	\caption{Expected exclusion regions at 95\% confidence level for mis-aligned 2HDM type I for different Higgs boson masses in the $H\to WW \to 4j$ decay channel with $Z\to \ell\ell$. The region including the alignment limit is inaccessible by the search channel studied in this work.}
	\label{fig:excl1}
\end{figure}

\begin{figure}[thb]
	\centering
	\includegraphics[width=0.45\textwidth]{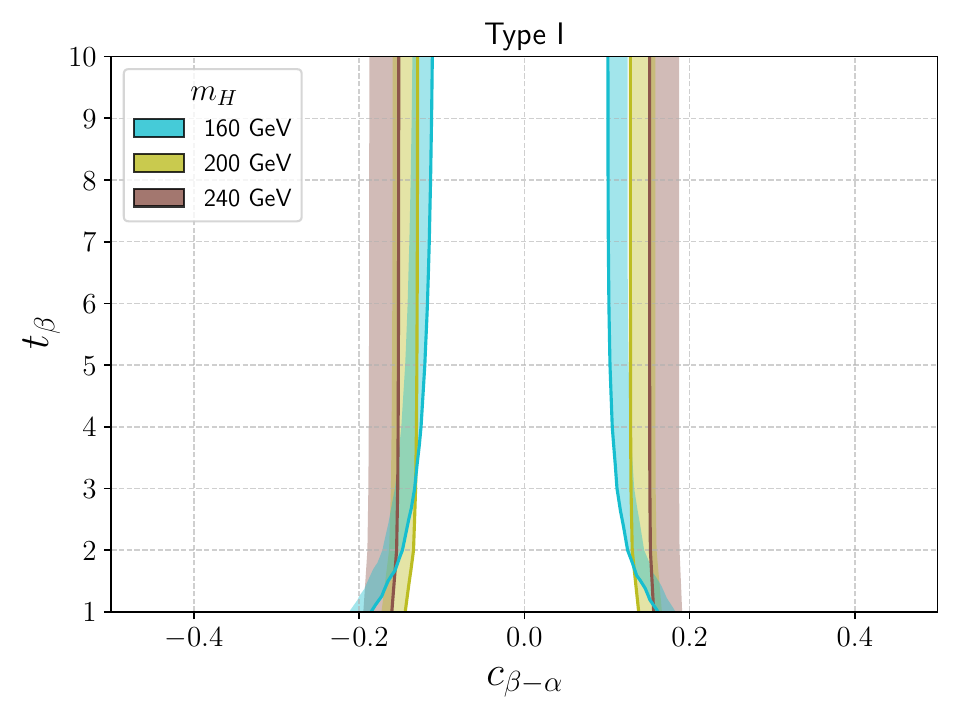}
	\caption{Expected exclusion regions at 95\% confidence level for mis-aligned 2HDM type I for different Higgs boson masses in the $H\to WW \to 4j$ decay channel with $Z\to \nu \nu$.}
	\label{fig:excl2}
\end{figure}

\begin{figure}[thb]
	\centering
	\includegraphics[width=0.45\textwidth]{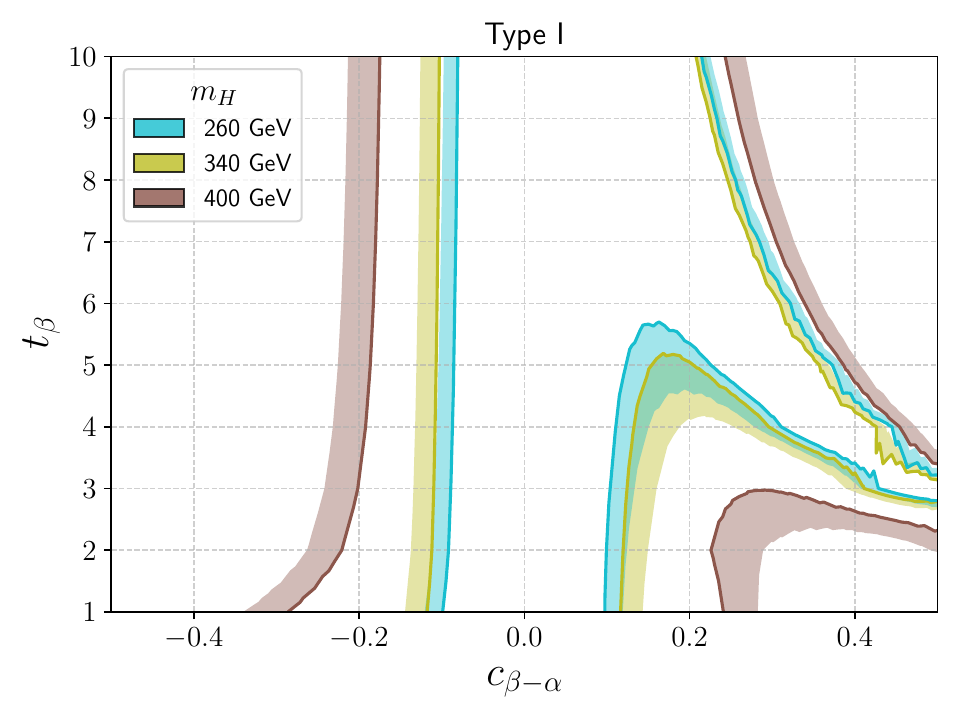}
	\caption{Expected exclusion regions at 95\% confidence level for mis-aligned 2HDM type I for different Higgs boson masses in the $H\to hh \to 4b$ decay channel with $Z\to \ell\ell$.}
	\label{fig:excl3}
\end{figure}

\begin{figure}[thb]
	\centering
	\includegraphics[width=0.45\textwidth]{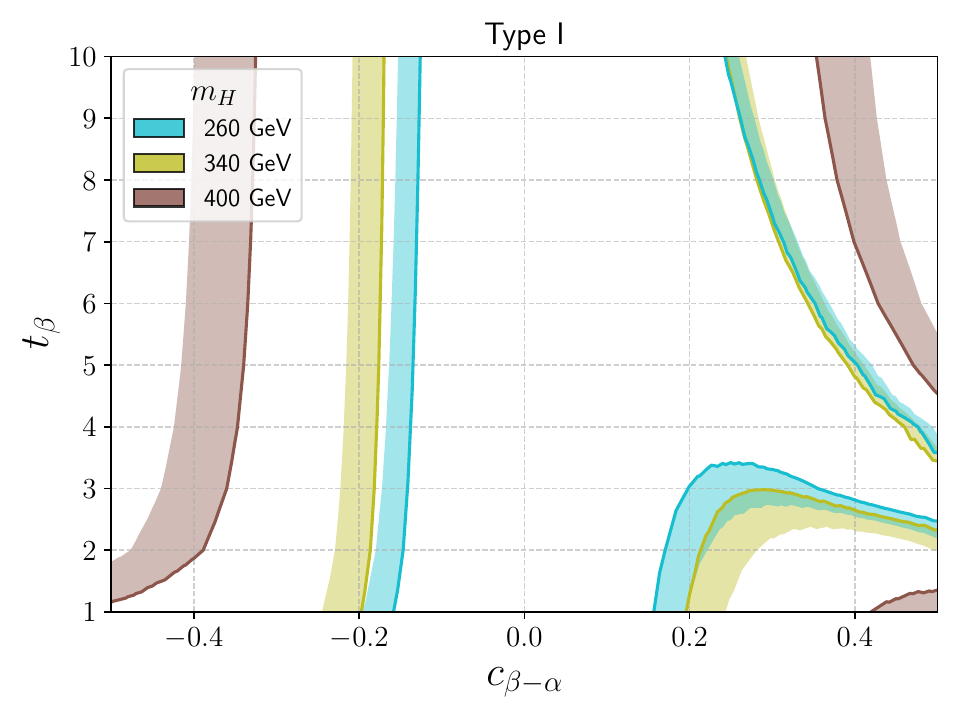}
	\caption{Expected exclusion regions at 95\% confidence level for mis-aligned 2HDM type I for different Higgs boson masses in the $H\to hh \to 4b$ decay channel with $Z\to \nu \nu$.}
	\label{fig:excl4}
\end{figure}

The resulting exclusion regions are shown in
\Cref{fig:excl1,fig:excl2,fig:excl3,fig:excl4} for selected Higgs boson
masses. Detector effects are modeled using the Delphes detector simulation,
including realistic object reconstruction efficiencies and
energy/momentum resolutions. No additional systematic uncertainties are
included, and the quoted exclusion limits therefore correspond to the expected
statistical sensitivity.

\section{Discussion and Conclusions}

A detailed study of heavy neutral Higgs boson production and
decay in the Type-I 2HDM was presented, emphasizing phenomenologically viable scenarios with small but finite departures from the alignment limit. After imposing
theoretical consistency conditions together with flavor, electroweak precision,
and Higgs signal-strength constraints, regions of parameter space are identified
where the heavy CP-even Higgs exhibits enhanced bosonic signatures while
fermionic modes are suppressed.

Our analysis shows that the Higgsstrahlung process $e^+ e^- \to ZH$ in the mis-aligned scenario at a future $e^+ e^-$ collider operating at $\sqrt{s} = 500$~GeV, combined
with the bosonic decay channels $H\to WW$ and $H\to hh$, can yield sizable
$\sigma\times\mathrm{BR}$ rates for moderate values of $\tb$. 

It is instructive to compare the present results with previous model-independent Higgsstrahlung studies based on the recoil-mass technique \cite{ZS}. Such analyses typically report exclusion limits on the reduced production cross section, $\sigma/\sigma_{\mathrm{SM}}$, which in the Type-I 2HDM is approximately related at tree level to the alignment parameter through
\begin{equation}
	\frac{\sigma(e^+e^-\to ZH)}{\sigma_{\mathrm{SM}}}\simeq c_{\beta-\alpha}^{,2}.
\end{equation}
For heavy scalar masses around $200~\mathrm{GeV}$, recoil-mass studies achieve sensitivities corresponding to $\sigma/\sigma_{\mathrm{SM}}\sim10^{-2}$, i.e. $|c_{\beta-\alpha}|\sim0.1$ \cite{ZS}, while the sensitivity deteriorates for larger masses as the Higgsstrahlung production cross section decreases near the kinematic threshold. 

The present analysis, based on direct reconstruction of the heavy Higgs boson in the $H\to WW$ and $H\to hh$ decay modes, demonstrates sensitivity to deviations from the alignment limit over the considered benchmark mass range. In the low-mass region, the expected $95\%$ confidence level exclusion reaches $|\cos(\beta-\alpha)|\gtrsim0.1$ for $m_H=160$~GeV and gradually weakens to approximately $|\cos(\beta-\alpha)|\gtrsim0.2$ at $m_H=240$~GeV. A comparable sensitivity is obtained for higher Higgs boson masses at low $\tan\beta$, while the exclusion becomes progressively weaker for larger $\tan\beta$ owing to the reduced signal production rate.
\section{Acknowledgements}
The computing facilities provided by Shiraz University, which were essential for this work, are gratefully acknowledged. 


%

\end{document}